%% file: msm_main.tex
\input{preamble.tex}

\journal{Computer Physics Communications}

\begin{document}

\begin{frontmatter}

  
  
  \title{msmJAX: Fast and Differentiable Electrostatics on the GPU in Python}
  
  
  \author[a]{Florian Buchner}
  \author[a]{Johannes Schörghuber}
  \author[a]{Nico Unglert}
  \author[b]{Jes\'us Carrete}
  \author[a]{Georg K. H. Madsen\corref{author}}
  
  \cortext[author] {Corresponding author.\\\textit{E-mail address:} georg.madsen@tuwien.ac.at}
  \address[a]{Institute of Materials Chemistry, TU Wien, A-1060 Vienna, Austria}
  \address[b]{Instituto de Nanociencia y Materiales de Aragón (INMA), CSIC-Universidad de Zaragoza, 50009 Zaragoza, Spain}
  
  \begin{abstract}
  
  
  
  

  We present \programnametext{}, a Python package implementing the multilevel summation method with B-spline interpolation, a linear-scaling algorithm for efficiently evaluating electrostatic and other long-range interactions in particle-based simulations.
  Built on the \formatprogramtitle{JAX} framework,  \programnametext{} integrates naturally with the machine-learning methods that are  transforming chemistry and materials science, while also serving as a powerful tool in its own right. It combines high performance with Python's accessibility, offers easy deployment on GPUs, and supports automatic differentiation.
  We outline the modular design of \programnametext{}, enabling users to adapt or extend the code, and present benchmarks and examples, including a verification of linear scaling, and demonstrations of its stability in molecular-dynamics  simulations.
  \\

     \end{abstract}
  \end{frontmatter}

\section{Introduction}
\label{sec_intro}

Evaluating pairwise long-range interactions, like electrostatic and dispersion interactions, that cannot be truncated without making significant errors is a ubiquitous task in atomistic simulations of matter.
Despite these interactions being described by simple expressions, their computation is demanding in practice as the cost of a direct evaluation scales as $\mathcal{O}(N^2)$ with the number of particles.
Another, more subtle, issue arises for periodic systems, where the electrostatic energy, a priori, is  only conditionally convergent, i.e., depends on the summation order.\cite{BGM+13}

Various schemes, that we will refer to as fast electrostatics algorithms, have been developed to address these challenges.
The classic way of assigning a physically meaningful value to the conditionally convergent periodic lattice energy is Ewald summation,\cite{Ewa21} which splits the Coulomb potential into a rapidly decaying part evaluated by summing over particles in real space and a long-range but smooth part evaluated by summing over reciprocal-lattice vectors.
More commonly used than Ewald's original method are several mesh-based modifications thereof that improve the computational efficiency by allowing the fast Fourier transform (FFT) to be used for the reciprocal-space part.\cite{DH98,DH98a}
These algorithms, which include particle–particle particle–mesh (P3M),\cite{p3m1988} particle-mesh Ewald (PME)\cite{pme1993}, and smooth particle-mesh Ewald (SPME),\cite{spme1995} achieve a scaling of $\mathcal{O}(N \log(N))$.

Despite the success of Ewald-based methods, their reliance on periodicity as well as issues with the parallel scalability of FFTs have sustained interest in alternatives, most notably the fast multipole method (FMM)\cite{GR87,CGR88,SYTH19} and the multilevel summation method (MSM).\cite{BL90,BV98,Har06,HWP+15,HWX+16}
Both FMM and MSM achieve $\mathcal{O}(N)$ scaling through a hierarchical separation of length scales.
Compared to the famous FMM, that has even been named one of the top 10 algorithms of the twentieth century,\cite{DS00} the MSM is less well known, but it has since been developed into a viable competitor, with implementations readily available in the \formatprogramtitle{NAMD}\cite{PHM+20} and \formatprogramtitle{LAMMPS}\cite{TAB+22} molecular-dynamics (MD) packages.
The MSM produces continuous energies and forces, which together with an appealing conceptual simplicity, natural handling of mixed periodicity, and straightforward extension to interactions other than electrostatics,\cite{TSBI14} makes it the method of choice for this work.
While our focus is on the pure MSM for now, we note that the distinctions made above are in fact not as absolute as they appear, with MSM-Ewald hybrid approaches having been proposed as well.\cite{KHS21}

Importantly, the need for fast electrostatics algorithms is not limited to evaluating the energy of fixed point charges in classical force fields, as mathematically equivalent or closely related problems also arise when long-range effects are to be incorporated into machine-learned interatomic potentials (MLIPs).\cite{KFG+21a,SWK+22,SPK+24,GC19,FKK24,KGGG23,Che25,FCMU24}
For the application with MLIPs, however, existing implementations of fast electrostatics algorithms largely fall short because they tend to be tightly integrated into monolithic MD packages, while modern MLIPs are built on machine-learning frameworks like \formatprogramtitle{TensorFlow},\cite{tensorflow2015-whitepaper} \formatprogramtitle{PyTorch}\cite{pytorch2019} or \formatprogramtitle{JAX}.\cite{jax2018github}

Research on long-range MLIPs could thus greatly benefit from the availability of fast electrostatics implementations in these frameworks.
In fact, their utility would go beyond MLIPs, because the underlying ML frameworks are also suitable as general-purpose scientific-computing libraries given that they offer high performance, graphics processing unit (GPU) support, and automatic differentiation, all while being programmable through a high-level Python interface.
Indeed, this problem has been recognized before, by the authors of the \formatprogramtitle{JAX MD} package,\cite{jaxmd2020} release v0.2.25 of which added\cite{jax-md_v0.2.25} code for Ewald summation and PME, as well as by Loche et al.\cite{LHH+25} who recently made Ewald, PME and P3M implementations available through the \formatprogramtitle{torch-pme} and \formatprogramtitle{jax-pme} packages.
Another recent example is the \formatprogramtitle{jaxFMM} package.\cite{jaxfmm2025}

We added to this toolbox of ML-framework-compatible fast electrostatics implementations by developing \programnametext, a Python package containing a \formatprogramtitle{JAX} implementation of the MSM, in particular the improved version of the algorithm from Ref.~\citenum{HWX+16} that employs B-spline interpolation.
Our code achieves linear scaling in the computation of electrostatic energies, forces, charge gradients, and stress tensors, with the latter three obtained via automatic differentiation of the energy. 
On a technical level, \programnametext{}  supports general parallelepipeds as unit cells, works in one, two and three dimensions, and handles periodic, nonperiodic, and mixed boundary conditions. Its standalone nature, modular design, accessibility through Python, and functional coding style that makes the underlying mathematics explicit, make \programnametext{} both a pedagogically valuable entry point to MSM, and a flexible framework for prototyping of extensions or modifications: an example would be hybrid approaches similar to Ref.~\citenum{KHS21}.

\section{Fundamentals of the multilevel summation method}
\label{sec_fundamentals_of_msm}

The MSM has been described in detail elsewhere.\cite{Har06,HWP+15,HWX+16}
To keep the presentation self-contained, we provide a brief overview of the formalism, using language and notation that mirror our code structure.
For the most part, our discussion and notation follow that in Ref.~\citenum{HWX+16}.
This section is meant as a high-level overview of the MSM, with more details discussed later alongside the specifics of their implementation, from \cref{ssec_coulomb_kernel_splitting} onward.

The MSM is an algorithm for the efficient approximate evaluation of sums, interpretable as a potential energy, of pairwise interactions of the form
\begin{equation}\label{eq:generic_pair_potential_sum}
  U
    = \frac{1}{2} \sumParticlesI \sumParticlesJExclI q_\particleIndexI q_\particleIndexJ k(r_{\particleIndexI \particleIndexJ})
  \, .
\end{equation}
Here, $q_\particleIndexI$ and $q_\particleIndexJ$ denote charges (or some more general kind of source strengths) associated with particles labeled $\particleIndexI$ and $\particleIndexJ$, respectively, $r_{\particleIndexI \particleIndexJ} = \lVert \posvec_\particleIndexI - \posvec_\particleIndexJ \rVert_2$ is their distance, and $k(r)$ is a two-body interaction potential that depends on a scalar distance argument alone.

\begin{figure*}
  \centering
  \includegraphics[width=.95\textwidth]{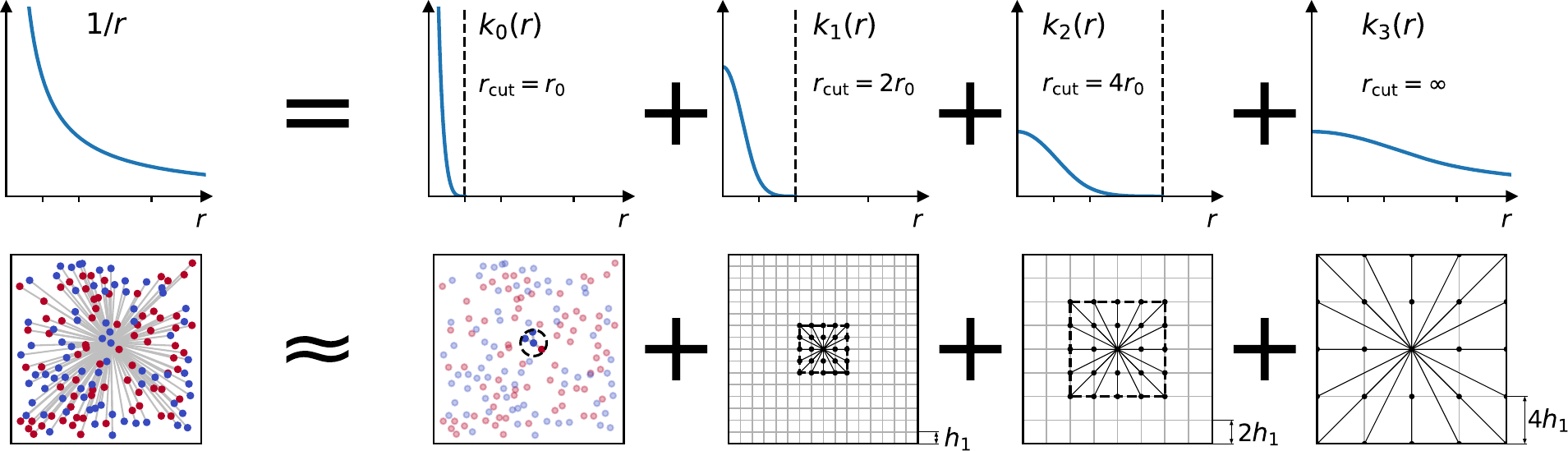}
  \caption{
    \label{fig:msm_basics}
    Basic principles of the multilevel summation method.
    Top row: splitting of a long-range potential (Coulomb potential $1/r$ in this example) into a sum of partial kernels with increasing cutoffs that also get smoother and smoother.
    Bottom row: schematic evaluation in both cases.
    Whereas exact evaluation of the original long-range potential (to the left of the approximately-equal sign) requires evaluating all pairwise interactions, the MSM (to the right of the approximately-equal sign) evaluates only the shortest-range kernel $k_0(r)$ directly and approximates the others from grids with increasing spacing.
    Levels increase from left to right, red and blue points symbolize particles of opposite charge.
  }
\end{figure*}

The main conceptual step is to split the interaction potential into a sum of \emph{partial kernels},
\begin{equation}\label{eq:kernel_splitting}
  k(r) = \sum_{\levelIndex=0}^L k_\levelIndex(r) = k_0(r) + k_1(r) + \dots + k_{L-1}(r) + k_L(r) \, ,
\end{equation}
where the indices $l$ refer to the \emph{levels} that give the method its name. The splitting is illustrated schematically in the top part of \cref{fig:msm_basics}.
With the exception of $k_L(r)$, which retains any infinite-range tail of the original $k(r)$, all partial kernels $k_{\levelIndex}(r)$ have a finite range.
As \levelIndex{} increases, the $k_\levelIndex(r)$ also get increasingly smooth and 
their cut-off radii $r_{\mathrm{cut}}^{(\levelIndex)}$ increase with $\levelIndex$, typically doubling from one level to the next,
\begin{equation}
  r_{\mathrm{cut}}^{(\levelIndex)} =
    \begin{cases}
      2^\levelIndex \levelZeroCutoff & \levelIndex = 0, \dots, L - 1 \, , \\
      \infty            & \levelIndex = L \, .
    \end{cases}
\end{equation}

With the kernel splitting, the energy \cref{eq:generic_pair_potential_sum} is expressed as a sum of two contributions,
\begin{equation}\label{eq:U_contributions}
  U = \Uzero + \Uoneplus \, ,
\end{equation}
where $\Uzero$ is a short-range contribution
\begin{equation}\label{eq:U_zero}
  \Uzero
    = \frac{1}{2} \sumParticlesI \sumParticlesJExclI q_\particleIndexI q_\particleIndexJ k_0(r_{\particleIndexI \particleIndexJ})
      - \frac{1}{2} \sumParticlesI q_\particleIndexI^2 \sumLevelsOnePlus k_\levelIndex(0)
  \, ,
\end{equation}
and $\Uoneplus$ is a long-range contribution
\begin{equation}\label{eq:U_oneplus}
  \Uoneplus
    =
      \frac{1}{2} \sumLevelsOnePlus \sumParticlesI \sumParticlesJ
      q_\particleIndexI q_\particleIndexJ
      k_\levelIndex(r_{\particleIndexI \particleIndexJ})
  \, .
\end{equation}
One subtlety to note about this formulation is that, to make subsequent manipulations more convenient, the constraint $\particleIndexI \neq \particleIndexJ$ has been dropped in \cref{eq:U_oneplus}.
This amounts to an unphysical self interaction, to correct for which the second term in \cref{eq:U_zero} has been introduced.
In doing so, we have assumed that any singularity of $k(r)$ at $r=0$ is contained in $k_0(r)$ alone, so that the $k_\levelIndex(0)$ for $\levelIndex = 1, \dots, L$ can be safely evaluated.

\begin{figure}[htb]
  \centering
  \includegraphics{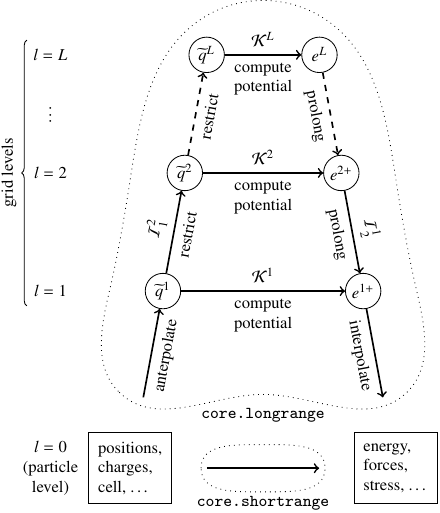}
  \caption{
    \label{fig:ladder_diagram}
    Abstract schematic of algorithm steps.
    The separation into short-range (evaluated directly) and long-range parts (approximated using grids) is highlighted, along with the corresponding modules of \programnametext.
    Rectangular boxes on the bottom left and right represent inputs and outputs, respectively, while round nodes in the upper part represent intermediate results defined on grids at different levels.
    The tapering of the ``ladder'' towards the top symbolizes the coarsening of subsequent grids.
    Arrows indicate the application of linear operators, and where two arrows end at the same node, their results are added together.
  }
\end{figure}

Up to this point, nothing more than rewriting has been done.
Next, the central approximation of the MSM is made, which concerns the long-range energy contribution \Uoneplus. As illustrated schematically \cref{fig:msm_basics}, the algorithm introduces real-space grids at all levels $\levelIndex \geq 1$ and interpolates the partial kernels $k_\levelIndex(r)$ in \cref{eq:U_oneplus} from these grids to approximate the long-range term \Uoneplus.
\cref{fig:ladder_diagram} illustrates the steps of the MSM in a way commonly found in the literature.\cite{Har06,HWP+15,HWX+16}
The exact evaluation of the short-range energy contribution \Uzero{} is represented by the lower part, whereas the approximate evaluation of \Uoneplus{} is represented by the upper part.

The approximation of the partial kernels for $\levelIndex \geq 1$ by their interpolants can be formally written, in three dimensions, as
\begin{equation}\label{eq:kernel_interpolation}
  \begin{gathered}
    k_l(\lVert \posvec - \posvec' \rVert_2)
      \approx \sum_\gridpointIndexM \sum_\gridpointIndexN
        \varphi^\levelIndex_\gridpointIndexM(\posvec)
        \operatorK^\levelIndex_{\gridpointIndexM,\gridpointIndexN}
        \varphi^\levelIndex_\gridpointIndexN(\posvec')
      \, , \ \levelIndex \geq 1 \, , \\
      \varphi^\levelIndex_\gridpointIndexM(\mathbf{r})
        = \varphi^{\levelIndex}_{m_x}(x) \, \varphi^{\levelIndex}_{m_y}(y) \, \varphi^{\levelIndex}_{m_z}(z) \, ,
  \end{gathered}
\end{equation}
where $\gridpointIndexM$, $\gridpointIndexN$ denote indices of grid points, $\varphi^\levelIndex_\gridpointIndexM(\mathbf{r})$ is a basis function centered on point $\gridpointIndexM$ of the level-\levelIndex{} grid, and $\operatorK^\levelIndex_{\gridpointIndexM,\gridpointIndexN}$ is a matrix of coefficients.
Inserting into \cref{eq:U_oneplus}, the long-range contribution becomes
\begin{equation}\label{eq:U_oneplus_grid_approx}
  \Uoneplus
    \approx
      \frac{1}{2} \sumLevelsOnePlus \sumGridpointsM \sumGridpointsN
      \gridCharge^{\levelIndex}_{\gridpointIndexM}
      \operatorK ^{\levelIndex}_{\gridpointIndexM , \gridpointIndexN}
      \gridCharge^{\levelIndex}_{\gridpointIndexN}
      \\
    \, ,
\end{equation}
where we have introduced the so-called \emph{grid charge},
\begin{equation}\label{eq:grid_charge}
  \gridCharge^\levelIndex_\gridpointIndexM
    =
      \sumParticlesI q_{\particleIndexI} \varphi^{\levelIndex}_{\gridpointIndexM}(\posvec_{\particleIndexI})
  \, ,
\end{equation}
defined at each grid point $\gridpointIndexM$.
The operation that computes it is referred to as an \emph{anterpolation} (a term introduced in Ref.~\citenum{Bra91} to mean the adjoint of interpolation) and corresponds to the lower left arrow in \cref{fig:ladder_diagram}, that maps from the particle level onto the grid level.
There is a formal analogy between \cref{eq:U_oneplus_grid_approx} and the exact expression \cref{eq:U_oneplus} that it approximates, only the contractions are performed over indices of grid points instead of particles.
For improved readability, we will often use a compact operator-vector notation that suppresses the grid point indices, in which \cref{eq:U_oneplus_grid_approx} becomes
\begin{equation}\label{eq:U_oneplus_grid_approx_matrix_vector}
    \Uoneplus
      \approx
        \frac{1}{2} \sumLevelsOnePlus
        (\formalVector{\gridCharge}^{\levelIndex})\transpose
        \formalMatrix{\operatorK}^{\levelIndex}
        \formalVector{\gridCharge}^{\levelIndex}
      \, .
\end{equation}

Another detail of crucial practical importance is that, since $k_{\levelIndex}(\lVert \posvec - \posvec' \rVert_2)$ depends only on the difference between the arguments $\posvec$ and $\posvec'$, the effective dimension of \operatorK{} is only half of what it would be in the fully general case.
In other words, the action of $\operatorK^{\levelIndex}$ on $\gridCharge^{\levelIndex}$ reduces from a general matrix-vector multiplication to a convolution with a stencil $\kernelStencil^{\levelIndex}$,
\begin{equation}\label{eq:action_of_K_as_convolution}
  \left( \formalMatrix{\operatorK}^{\levelIndex} \formalVector{\gridCharge}^{\levelIndex} \right)_{\gridpointIndexM}
  =
  \sumGridpointsN \operatorK^{\levelIndex}_{\gridpointIndexM, \gridpointIndexN} \gridCharge^{\levelIndex}_{\gridpointIndexN}
  =
  \sumGridpointsN \kernelStencil^{\levelIndex}_{\gridpointIndexM - \gridpointIndexN} \gridCharge^{\levelIndex}_{\gridpointIndexN}
  \, .
\end{equation}
Note that our notation differs from that of Ref.~\citenum{HWX+16} in that prefactors $(2^\levelIndex \levelZeroCutoff)^{-1}$, if present, are absorbed into the $\formalMatrix{\kernelStencil}^{\levelIndex}$.

Since the functions $k_\levelIndex(r)$ supposed to be interpolated were previously constructed to be smoother the higher \levelIndex{} is, correspondingly coarser grids can be used for higher levels \levelIndex, with obvious benefits in terms of computational cost.
Typically, the grid spacings double from one level to the next,
\begin{equation}\label{eq:grid_spacing_doubling}
  h_{\levelIndex} = 2^{\levelIndex - 1} h_1 \, ,
\end{equation}
in tandem with the potential cutoffs, as illustrated in \cref{fig:msm_basics}.
Roughly speaking, the efficient scaling of the MSM comes from the fact that the computation of the grid charge \cref{eq:grid_charge} scales linearly with $N$, and that the matrices $\formalMatrix{\operatorK}^\levelIndex$ in \cref{eq:U_oneplus_grid_approx} remain sparse even as the cutoff radii of the higher-level $k_l(r)$ grow very large, since the grid spacing increases along with the cutoff.

The efficiency of the MSM furthermore relies on the introduction of \emph{translation operators} $\formalMatrix{\operatorI}^{\levelIndex}_{\levelIndex-1}$ with the property $\gridCharge^\levelIndex = \formalMatrix{\operatorI}^{\levelIndex}_{\levelIndex-1} \gridCharge^{\levelIndex-1}$.
Depending on the exact choice of basis functions $\varphi^{\levelIndex}_{\gridpointIndexM}(\posvec)$ and the layout of grids at subsequent levels relative to each other, the equality may hold exactly or only approximately.
Repeatedly applying the translation operators allows building up all higher-level grid charges from the lowest-level one, $\gridCharge^\levelIndex = \formalMatrix{\operatorI}^{\levelIndex}_{\levelIndex-1} \dots \formalMatrix{\operatorI}^{3}_{2} \formalMatrix{\operatorI}^{2}_{1} \gridCharge^{1}$, so evaluation of \cref{eq:grid_charge} is now only required for $ l = 1 $.
The grid $\leftrightarrow$ grid operations effected by the translation operators are referred to as \emph{restriction} and \emph{prolongation}.
The term restriction is used for the operators $\formalMatrix{\operatorI}^{\levelIndex}_{\levelIndex-1}$ that transport charge from a lower (finer) to a higher (coarser) grid. They are represented by upward arrows between round nodes on the left of \cref{fig:ladder_diagram}.
Closely related are the prolongation operators $\formalMatrix{\operatorI}^{\levelIndex-1}_{\levelIndex} \coloneqq ( \formalMatrix{\operatorI}^{\levelIndex}_{\levelIndex-1} )\transpose$ that transport electrostatic potentials originally calculated at a higher (coarser) grid to a lower (finer) grid.
These are represented in \cref{fig:ladder_diagram} by downward arrows between round nodes on the right.

The third type of grid $\leftrightarrow$ grid operators, $\formalMatrix{\operatorK}^\levelIndex$, compute (coefficients for interpolating) the electrostatic potentials due to the partial kernels $k_\levelIndex(r)$ at the corresponding grid levels, and are represented in \cref{fig:ladder_diagram} by the horizontal left-to-right arrows in the upper part.
Combining the three types of grid $\leftrightarrow$ grid operators leads to rewriting \cref{eq:U_oneplus_grid_approx} in a nested fashion as
\begin{equation}\label{eq:u_oneplus_nested_eval}
  \begin{gathered}
  \Uoneplus
    \approx \frac{1}{2} \left( \formalVector{\gridCharge}^1 \right)\transpose \formalVector{e}^{1+}
    \, , \\
  \formalVector{e}^{1+} \coloneqq \Big[
    \formalMatrix{\operatorK}^1 + \formalMatrix{\operatorI}^1_2 \Big(
      \formalMatrix{\operatorK}^2 + \formalMatrix{\operatorI}^2_3 \big(
        \dots
      \big) \formalMatrix{\operatorI}^3_2
    \Big) \formalMatrix{\operatorI}^2_1
  \Big] \formalVector{\gridCharge}^1
  \, .
  \end{gathered}
\end{equation}
Here, we have highlighted a useful intermediate result, $\formalVector{e}^{1+}$, that, for practical purposes, is often beneficial to store, and work in terms of.

The final step of the computation of \Uoneplus{}, the contraction of $\formalVector{e}^{1+}$ with $\formalVector{\gridCharge}^1$, involves evaluating the interpolant to the electrostatic potential at the particle positions $\posvec_\particleIndexI$, and the name \emph{interpolation} is conventionally given to this algorithm step in its entirety.
It is represented by the lower right arrow in \cref{fig:ladder_diagram} and can be thought of as going back from the grid to the particle level.
This becomes clearer when rewriting \cref{eq:u_oneplus_nested_eval} by re-inserting the definition \cref{eq:grid_charge} of $\formalVector{\gridCharge}$,
\begin{equation}\label{eq:uoneplus_interpolation}
  \Uoneplus \approx
      \frac{1}{2}
      \left( \formalVector{\gridCharge}^1 \right)\transpose \formalVector{e}^{1+}
    = \frac{1}{2}
      \sumParticlesI q_\particleIndexI
      \sum_{\gridpointIndexM} e^{1+}_\gridpointIndexM \varphi^1_\gridpointIndexM (\posvec_\particleIndexI)
    \, ,
\end{equation}
which reveals the meaning of $\formalVector{e}^{1+}$ as a set of coefficients from which the electrostatic potential can be interpolated.
While explicitly re-introducing the basis functions instead of working with $\formalVector{\gridCharge}^1$ directly may seem superfluous, the formulation in \cref{eq:uoneplus_interpolation} will prove useful for computing derivatives of \Uoneplus{}.
This will be elaborated on in \cref{sssec_impl_core_longrange}.

We can thus summarize the MSM algorithm as follows:
On top of a particle $\leftrightarrow$ particle part that performs a standard evaluation of a short-range pair potential, it consists of two particle $\leftrightarrow$ grid operations, anterpolation and interpolation, moving between the particle and the grid level, and a set of grid $\leftrightarrow$ grid operations in between.
Whereas the anterpolation step transforms the particle charges to a grid-based representation $\gridCharge^1$, the interpolation step reconstructs the electrostatic potential at particle positions from its grid representation $\formalVector{e}^{1+}$.
The grid $\leftrightarrow$ grid operations, corresponding to the application of the nested operator sum inside the brackets in \cref{eq:u_oneplus_nested_eval}, compute $\formalVector{e}^{1+}$ from $\gridCharge^1$.

\section{Implementation details}
\label{sec_implementation_details}

We implemented \programnametext{} in a mostly functional coding style, where the outputs of the model setup are side-effect-free and free-floating functions that evaluate the electrostatic properties, and the main inputs to the setup are themselves lower-level functions.
This design, that promotes functions to fundamental elements, is in line with the philosophy of \formatprogramtitle{JAX}, where key features\textemdash like just-in-time compilation via \formatcode{jax.jit}, automatic differentiation via \formatcode{jax.grad}, and vectorization via \formatcode{jax.vmap}\textemdash take the form of function transformations that act on functions and return modified versions thereof. At the same time, the approach lends itself naturally to a modular implementation of the MSM. What distinguishes different varieties of the algorithm is often a different choice of a particular function, like which interaction potential $k(r)$ to evaluate or which basis functions $\varphi^{\levelIndex}_{\gridpointIndexM} (\posvec)$ are used for interpolation.

\subsection{Package structure}
\label{ssec_package_structure}

\begin{figure}
  \centering
  \includegraphics[width=\columnwidth]{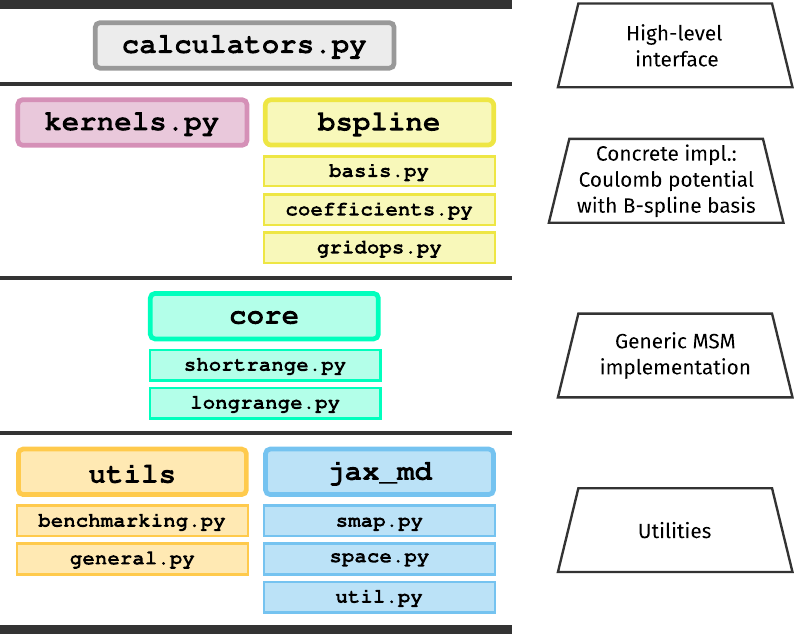}
  \caption{
    \label{fig:package_structure}
    Package structure of \programnametext.
    The larger boxes with rounded edges represent either top-level module files (\formatcode{.py} extension) or subpackages (no extension).
    Modules contained in these subpackages are in turn represented by the smaller rectangular boxes of the same color below them.
    The arrangement of components from bottom to top reflects a conceptual hierarchy of increasingly higher-level functionalities.
  }
\end{figure}

Out of the box, \programnametext{} contains a ready-to-use implementation of the MSM specifically for Coulomb potentials (i.e., that evaluates $k(r) = 1/r$), and that employs B-splines for the basis functions $\varphi^{\levelIndex}_{\gridpointIndexM}(\posvec)$.
A high-level interface to this implementation is provided by the \formatcode{\programnamecode.calculators} module, placed at the top of the package structure shown in \cref{fig:package_structure}.
For many applications and users, this high-level interface may be all that is required.

Internally, this concrete implementation is kept separate from a lower-level core algorithm designed to be sufficiently generic to be reusable when adapting \programnametext{} to other interaction potentials or interpolation schemes.
This lower-level algorithm is contained in \formatcode{\programnamecode.core}, further divided into independent \formatcode{shortrange} and \formatcode{longrange} modules, while the extra components necessary for the concrete implementation are in \formatcode{\programnamecode.kernels} and \formatcode{\programnamecode.bspline}.

The \formatcode{\programnamecode.utils} subpackage contains miscellaneous utilities either underpinning higher-level components or providing convenience functionality that does not strictly tie in with the hierarchy of the main code.
Lastly, the \formatcode{\programnamecode.jax\_md} subpackage contains code from the \formatprogramtitle{JAX MD}\cite{jaxmd2020} library that we use in the short-range evaluation (see \cref{sssec_impl_core_shortrange} for details) and chose to include with \programnametext{} in order to reduce dependencies, given that we use only a very small subset of the code of \formatprogramtitle{JAX MD}.

\subsection{MSM core algorithm}
\label{ssec_impl_core}

\begin{figure}[htb]
  \centering
  \includegraphics[width=\columnwidth]{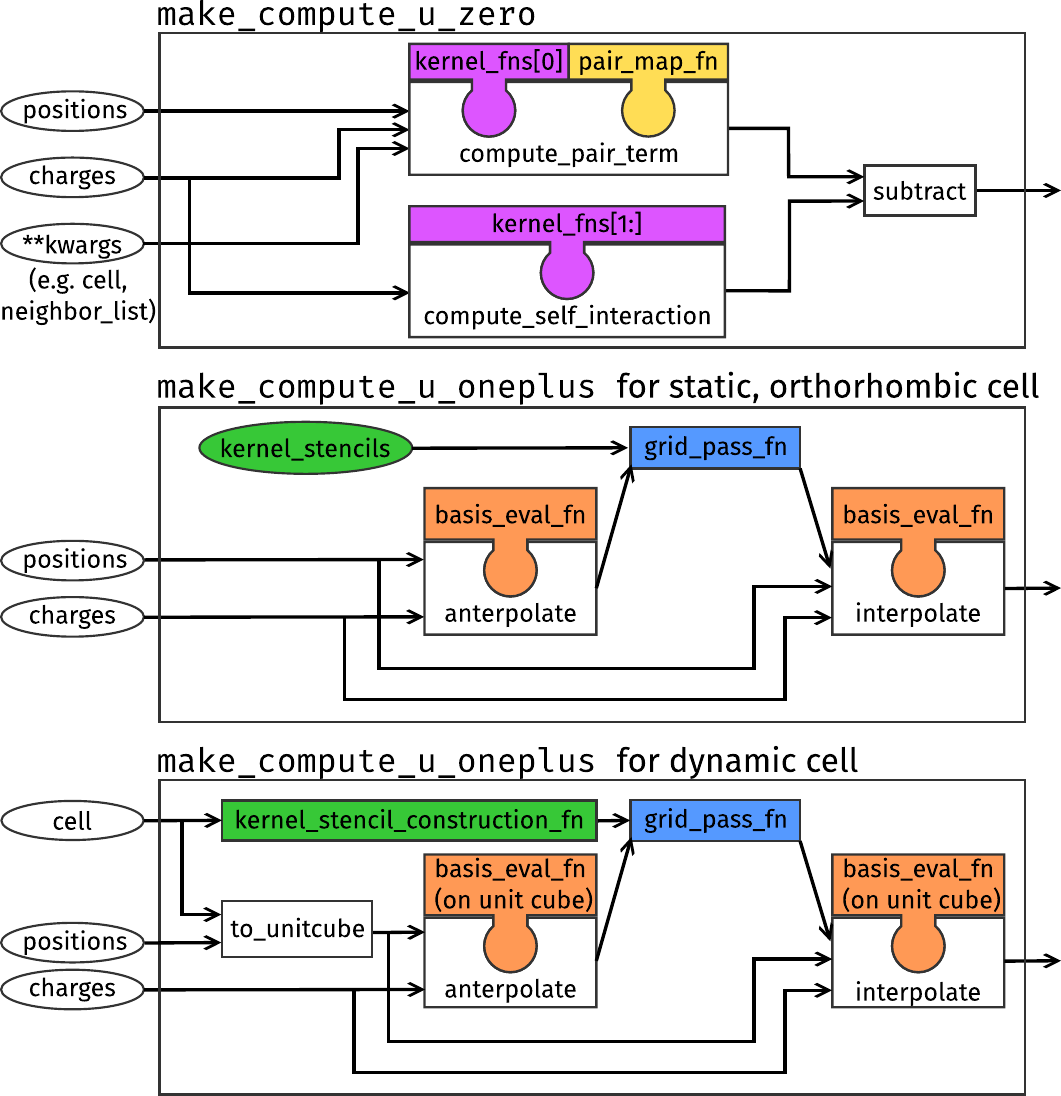}
  \caption{
    \label{fig:core_modularity}
    Simplified schematic of how the setup functions \formatcode{make\_compute\_u\_zero} and \formatcode{make\_compute\_u\_oneplus} provided in \formatcode{\programnamecode.core} construct evaluation functions for \Uzero{} and \Uoneplus.
    The elements with a color fill are the inputs to the setup functions, in the choice of which lies the modularity of the design.
    Ellipses represent data and rectangles represent functions, with their inputs/outputs represented by incoming/outgoing arrows.
    In particular, the outermost enclosing rectangles correspond to the end-to-end evaluation functions that are constructed.
    Where blocks slot into other blocks in the style of jigsaw puzzle pieces, this emphasizes how one function internally serves as input to the construction of another, more complex one.
  }
\end{figure}

\subsubsection{Short-range part}
\label{sssec_impl_core_shortrange}

The \formatcode{core.shortrange} module contains functionality for the direct (i.e., without interpolation grids) evaluation of the short-range contribution \Uzero, given by \cref{eq:U_zero}.
This corresponds to the lower part of the algorithm schematic in \cref{fig:ladder_diagram}.

As its main convenience interface, the module provides a setup function \formatcode{make\_compute\_u\_zero} to construct an evaluation function for \Uzero.
This setup function takes two inputs, a sequence \formatcode{kernel\_fns} of one-argument kernel functions $k_\levelIndex(r)$ (one element for each level $\levelIndex$), and a function that we will call \formatcode{pair\_map\_fn}.
The \formatcode{pair\_map\_fn} is a higher-order function that transforms a pair potential (a function of a single scalar distance argument) into another function that computes the total system energy due to that pair potential from the set of all particle positions and charges, as well as optionally from arbitrary extra arguments handled via keywords.
Thus, while \formatcode{kernel\_fns} specifies \emph{which} interaction potential to evaluate, \formatcode{pair\_map\_fn} specifies \emph{how} to apply it to a system of particles.
The way these two inputs to \formatcode{make\_compute\_u\_zero} are used internally to modularly construct an evaluator for \Uzero{} is illustrated in the top panel of \cref{fig:core_modularity}.
The first element of \formatcode{kernel\_fns}, corresponding to the shortest-range kernel $k_0(r)$, together with \formatcode{pair\_map\_fn} is used to construct an evaluator for the pair term in \cref{eq:U_zero}.
The remaining elements of \formatcode{kernel\_fns}, corresponding to the higher-level kernels $k_\levelIndex(r)$, $\levelIndex = 1, \dots, L$, meanwhile enter into the calculation of the straightforward and inexpensive self-interaction term in \cref{eq:U_zero}.

The \formatcode{pair\_map\_fn} plays an important role in this design, for two reasons.
First, it is performance-relevant as it is responsible for the pairwise distance computation and mapped potential evaluation contained in \Uzero.
Second, all options relating to these pairwise operations, like the specification of periodic boundary conditions, or the choice to use a neighbor list for evaluation or not, are selected by supplying an appropriately defined \formatcode{pair\_map\_fn}.
In practice, users will not usually need to write their own \formatcode{pair\_map\_fn} implementation, as suitable ones for most use cases can be obtained using the \formatcode{make\_eval\_pair\_pot} and \formatcode{make\_eval\_pair\_pot\_neighborlist} functions provided in \formatcode{core.shortrange}.
These functions take a kernel function along with a number of options, most notably the periodicity along each axis,  as arguments, and return an evaluation function for the total system energy.
The code snippet below demonstrates their use in setting up a \formatcode{pair\_map\_fn} for use with \formatcode{make\_compute\_u\_zero}.
The example uses \formatcode{functools.partial} of the Python standard library to make the signature compatible as the \formatcode{pair\_map\_fn} is expected to take only one argument.
\begin{lstlisting}[language=Python,firstline=8]
pair_map_fn = functools.partial(
    make_eval_pair_pot,
    pbc=(True, True, False),  # fixing extra args
    cell_mode="triclinic",  # fixing extra args
)
compute_u_zero = make_compute_u_zero(
    kernel_fns, pair_map_fn
)
u_zero = compute_u_zero(
    positions, charges, cell=cell
)
\end{lstlisting}

The design is inspired by, and reuses some lower-level components from, the \formatprogramtitle{JAX MD}\cite{jaxmd2020} package.
More specifically, the previously introduced functions \formatcode{make\_eval\_pair\_pot} and \formatcode{make\_eval\_pair\_pot\_neighborlist} from \programnametext{} can be thought of as playing similar roles as \formatcode{smap.pair} and \formatcode{smap.pair\_neighbor\_list} in \formatprogramtitle{JAX MD}, only with the addition of particle charge arguments treated on an equal footing with positions, and with a slightly simplified interface more tailored to the concrete application.

\subsubsection{Long-range part}
\label{sssec_impl_core_longrange}
The \formatcode{core.longrange} module consists of generic code for evaluating the long-range contribution \Uoneplus{} (\cref{eq:U_oneplus}) using interpolation grids.
This corresponds to the upper part of \cref{fig:ladder_diagram}.
Analogously to the structure of the short-range implementation, the principal interface is a setup function named \formatcode{make\_compute\_u\_oneplus} that creates an evaluator for \Uoneplus.

The argument structure of this setup function reflects the distinction made in \cref{sec_fundamentals_of_msm} between particle $\leftrightarrow$ grid operations and grid $\leftrightarrow$ grid operations.
Its first input is a function \formatcode{basis\_eval\_fn} to evaluate the basis function $\varphi^{1}_{\gridpointIndexM}(\posvec)$ for a single particle.
This \formatcode{basis\_eval\_fn} is used in constructing the operations of the first kind (particle $\leftrightarrow$ grid), comprising anterpolation and interpolation, which both involve evaluation of the basis.
The second input to \formatcode{make\_compute\_u\_oneplus} is a function \formatcode{grid\_pass\_fn} that performs an entire end-to-end pass through all the operations of the second kind (grid $\leftrightarrow$ grid), i.e., going from the level-one grid charge $\formalVector{\gridCharge}^1$ to the accumulated electrostatic potential $\formalVector{e}^{1+}$.
The role played by these input functions as building blocks for the evaluation function for \Uoneplus{} is illustrated in the middle and bottom panels of \cref{fig:core_modularity}.

While this fact was not made explicit in \cref{sec_fundamentals_of_msm} for simplicity of notation, we assume finite support for the basis functions, limiting the terms that need to be summed over during the anterpolation and interpolation steps.
To make use of this fact, the setup function \formatcode{make\_compute\_u\_oneplus} expects the supplied \formatcode{basis\_eval\_fn} to return two outputs, the index set
\begin{equation}
  \mathcal{M}^{1}_{\particleIndexI} \coloneqq \{\gridpointIndexM : \varphi^{1}_{\gridpointIndexM}(\posvec_i) \neq 0 \}
\end{equation}
of all nearby grid points with non-zero basis function values around a particle's position, and the corresponding basis function values $\varphi^{1}_{\gridpointIndexM}(\posvec_i) \ \forall \ \gridpointIndexM \in \mathcal{M}^{1}_{\particleIndexI}$.
The anterpolation step then takes the form of a double loop (implemented in practice as a broadcast sum via \formatprogramtitle{JAX}'s \formatcode{.at[]} syntax), stated in \cref{alg:anterpolation_double_loop}, with the outer loop over particles and the inner one over nearby grid points.
\begin{algorithm}
  \caption{Formal double loop structure of the anterpolation implementation in \formatcode{make\_compute\_u\_oneplus}.}
  \label{alg:anterpolation_double_loop}
  \begin{algorithmic}
    \State Initialize all elements $\gridCharge^{1}_{\gridpointIndexM}$ to zero
    \For {$i \in \{1, \dots, N\}$}
      \For {$\gridpointIndexM \in \mathcal{M}^{1}_{\particleIndexI}$}
        \State $\gridCharge^{1}_{\gridpointIndexM} \gets \gridCharge^{1}_{\gridpointIndexM} + q_{\particleIndexI} \varphi^{1}_{\gridpointIndexM}(\posvec_{\particleIndexI})$
      \EndFor
    \EndFor
  \end{algorithmic}
\end{algorithm}
Similarly, the interpolation step is implemented making use of the finite support as
\begin{equation}\label{eq:uoneplus_interpolation_with_index_set}
  \Uoneplus
    \approx \frac{1}{2}
      \left( \formalVector{\gridCharge}^1 \right)\transpose \formalVector{e}^{1+}
    = \frac{1}{2}
      \sumParticlesI q_\particleIndexI
      \sum_{\gridpointIndexM \in \mathcal{M}^{1}_{\particleIndexI}} e^{1+}_\gridpointIndexM \varphi^1_\gridpointIndexM (\posvec_\particleIndexI)
  \, .
\end{equation}

The \formatcode{make\_compute\_u\_oneplus} wrapper does not take care of the internal logic of the grid pass on its own, i.e., the \formatcode{grid\_pass\_fn} supplied to it is expected to handle it in its entirety, in principle offering a high degree of flexibility.
Concrete implementations, however, will usually not need to include their own \formatcode{grid\_pass\_fn} written from scratch, but only functions for the restriction and prolongation steps, since the \formatcode{core.longrange} module provides an additional helper function \formatcode{make\_grid\_pass\_fn} that constructs an appropriate function from them according to the logic of \cref{eq:u_oneplus_nested_eval,fig:ladder_diagram}.

Our implementation comes with some optimizations to the way derivatives of \Uoneplus{} are computed.
The derivatives of the right-hand side of \cref{eq:uoneplus_interpolation_with_index_set} w.r.t.~particle positions and charges can be simplified to
\begin{equation}\label{eq:analytic_uoneplus_position_grad}
  \nabla_\particleIndexI \Uoneplus
    =
      q_{\particleIndexI} \sum_{\gridpointIndexM \in \mathcal{M}^{1}_{\particleIndexI}}
      e^{1+}_\gridpointIndexM
      \nabla \varphi^1_\gridpointIndexM (\posvec_\particleIndexI)
  \, ,
\end{equation}
\begin{equation}\label{eq:analytic_uoneplus_charge_grad}
  \frac{\partial \Uoneplus}{\partial q_{\particleIndexI}}
    =
      \sum_{\gridpointIndexM \in \mathcal{M}^{1}_{\particleIndexI}} e^{1+}_\gridpointIndexM \varphi^1_\gridpointIndexM (\posvec_\particleIndexI)
  \, ,
\end{equation}
from which it is apparent that the level-one accumulated potential $\formalVector{e}^{1+}$ can be reused between the calculation of \Uoneplus{} and its derivatives, and that no derivatives need to be propagated through it.
For the positions gradient \cref{eq:analytic_uoneplus_position_grad}, evaluating the derivative is only required for the basis functions $\varphi^1_\gridpointIndexM$, while the charge gradient \cref{eq:analytic_uoneplus_charge_grad} can even be fully expressed in terms of previously calculated quantities.

We implement these optimized derivatives by equipping the evaluation function for \Uoneplus{} returned by \formatcode{make\_compute\_u\_oneplus} with a custom Jacobian-vector product (JVP) operator, via \formatcode{jax.custom\_jvp}.
Inside the custom JVP, the gradients $\nabla \varphi^1_\gridpointIndexM (\posvec_\particleIndexI)$\textemdash whose functional form is unknown since the basis functions are themselves a user input to \formatcode{make\_compute\_u\_oneplus}\textemdash meanwhile continue to be evaluated by \formatprogramtitle{JAX}'s standard forward-mode automatic differentiation, combining the improved efficiency of the optimized derivative with the convenience and flexibility of automatic differentiation that works on arbitrary functions.
With the evaluation function decorated in such a way, all of \formatprogramtitle{JAX}'s automatic-differentiation operations called on it, or on any code wrapping it, by default make use of the optimizations with no additional implementation effort for users.
The performance gain is documented in \cref{sssec_benchmarks_custom_deriv}.

\newcommand{\posvecunitcube}{\ensuremath{\mathbf{s}}}

The \formatcode{core.longrange} module also provides the general framework for how to evaluate \Uoneplus{} in cells with non-orthogonal basis vectors.
We focus here on the interface of our implementation, and refer to Section 6.2.5 of Ref.~\citenum{Har06} for more mathematical background.
To make an approximation like \cref{eq:kernel_interpolation} applicable, the problem is first transformed to a rectangular coordinate system, in which the grid operations are then performed as usual.
We choose this new coordinate system to always be the unit cube, i.e., for basis vectors $\mathbf{a}$, $\mathbf{b}$, $\mathbf{c}$ of the original unit cell, the position vectors $\posvec$ are mapped to the new ones by $\posvecunitcube = [\mathbf{a}, \mathbf{b}, \mathbf{c}]^{-1} \posvec$.
That this transformation should be performed is signaled during setup by the presence of an optional \formatcode{transform\_mode} argument to \formatcode{make\_compute\_u\_oneplus}.
If this argument is given, users are expected to supply a \formatcode{basis\_eval\_fn} defined on the unit cube, i.e., it should act on $\posvecunitcube$.
Note that, when calculating forces or stresses in such situations via automatic differentiation of the energy w.r.t.~the untransformed positions or cell, no manual transformation back to Cartesian coordinates is required as this is automatically included as a consequence of differentiating through the transformation to the unit cube.

A related consideration is that if the cell volume or shape are allowed to change, like in constant-pressure MD simulations, the kernel coefficient stencils $\kernelStencil^{\levelIndex}$ need to be recomputed at each step.
This is because they depend on the values of the kernels $k_{\levelIndex}(\posvec)$ on a grid of distances congruent with the interpolation grids, which scale with the simulation cell.
This is handled via two additional, mutually exclusive, arguments to \formatcode{make\_compute\_u\_oneplus}, named \formatcode{kernel\_stencils} and \formatcode{kernel\_stencil\_construction\_fn}.
If the cell is fixed, the precomputed, constant, $\kernelStencil^{\levelIndex}$ can be supplied via the former argument, whereas for a dynamic cell, the latter argument is used.
It must be a function that dynamically computes all $\kernelStencil^{\levelIndex}$ from the current cell as input.

In \cref{fig:core_modularity} we contrast the case of a static and orthorhombic cell (middle panel), requiring neither transformation to the unit cube nor on-the-fly computation of the stencils $\kernelStencil^{\levelIndex}$, with that of a dynamic cell (bottom panel), requiring both.
While in the first case no cell argument to the top-level evaluation function for \Uoneplus{} is thus necessary, it is needed in the second, entering both into the transformation and stencil computation step.

\subsection{B-spline interpolation}
\label{ssec_bspline_interp}
The introduction of B-spline interpolation to the MSM in Ref.~\citenum{HWX+16}, on which \programnametext{} is based, marked a significant improvement of the algorithm.
In this section, we aim to highlight the aspects specific to this particular choice of interpolation scheme, and the corresponding components of \programnametext, all in the \formatcode{bspline} subpackage.
These would need replacement when switching to a different interpolation scheme.

\begin{figure}
  \centering
  \includegraphics[width=\columnwidth]{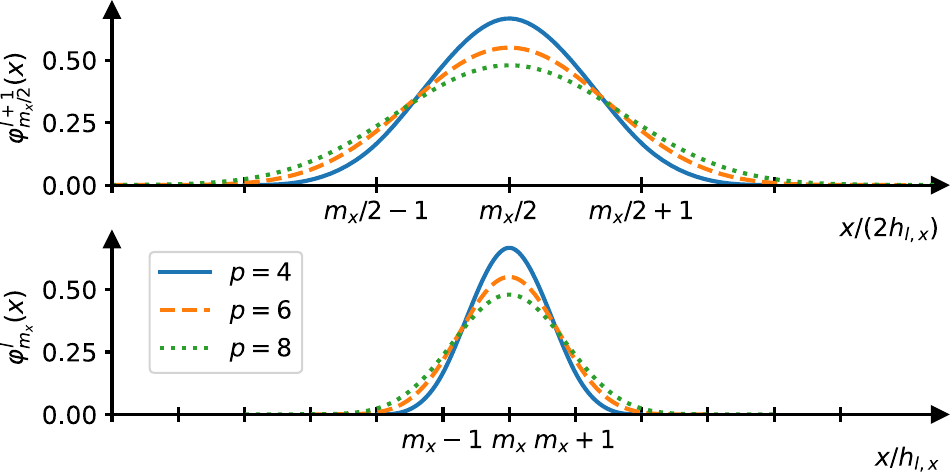}
  \caption{
    \label{fig:bspline_basis}
    B-spline basis functions at two successive grid levels $l$ (bottom) and $l+1$ (top), for different interpolation orders $p$.
  }
\end{figure}

First, the anterpolation step \cref{eq:grid_charge} requires evaluating the basis functions of the lowest-level ($\levelIndex = 1$) grid at the particle positions $\posvec_\particleIndexI$, i.e., calculating $\varphi^1_\gridpointIndexM (\posvec_\particleIndexI) = \varphi^{1}_{m_x}(x_\particleIndexI) \, \varphi^{1}_{m_y}(y_\particleIndexI) \, \varphi^{1}_{m_z}(z_\particleIndexI)$.
For the interpolation scheme proposed in Ref.~\citenum{HWX+16} and implemented in \programnametext{}, $\varphi^{1}_{m_x}$ is the cardinal B-spline basis function with knot spacing $h_{1,x}$ centered at the grid point $m_x$ (analogously for $y$, $z$).
Basis functions of different smoothness are distinguished by an interpolation-order parameter $p$ that is equal to the number of grid cells over which their support extends (see \cref{fig:bspline_basis}), and that is one greater than the polynomial degree of the pieces of the spline.
This usage of $p$ is in line with Ref.~\citenum{HWX+16}.
To evaluate the basis functions, we follow the recursion formula by de Boor,\cite{de_boor_bsplines_1972} implemented in the \formatcode{evaluate\_basis\_element} routine of the \formatcode{bspline.basis} module.
By fixing the interpolation order at runtime, the implementation is made compatible with the \formatcode{jax.jit} transformation.

Second, the restriction and prolongation operators, $\formalMatrix{\operatorI}^{\levelIndex}_{\levelIndex-1}$ and $\formalMatrix{\operatorI}^{\levelIndex}_{\levelIndex+1}$, need to be implemented for the chosen interpolation scheme.
Their action is defined by
\begin{equation}\label{eq:restriction_prolongation_explicit_formulas}
  \begin{split}
    \left( \formalMatrix{\operatorI}^{\levelIndex}_{\levelIndex-1} \gridCharge^{\levelIndex-1} \right)_{\gridpointIndexM}
    & = \sum_{\gridpointIndexN} J_{\gridpointIndexN - 2 \gridpointIndexM} \left( \gridCharge^{\levelIndex-1} \right)_{\gridpointIndexN} \, ,
    \\
    \left( \formalMatrix{\operatorI}^{\levelIndex}_{\levelIndex+1} e^{(\levelIndex+1)+} \right)_{\gridpointIndexM}
    & = \sum_{\gridpointIndexN} J_{\gridpointIndexM - 2 \gridpointIndexN} \left( e^{(\levelIndex+1)+} \right)_{\gridpointIndexN} \, ,
  \end{split}
\end{equation}
with $J_{\gridpointIndexN} = J_{n_x} J_{n_y} J_{n_z}$ containing universal coefficients that can be precomputed from known properties of the B-spline basis according to an uncomplicated formula (cf.~Ref.~\citenum{HWX+16} for more details).
Note that in the second line of \cref{eq:restriction_prolongation_explicit_formulas} we have corrected a likely typographic error contained in the formula for prolongation in section III.A.2.d of Ref.~\citenum{HWX+16}, where $\gridpointIndexM$ and $\gridpointIndexN$ are switched.
In \programnametext, the operators are implemented by \formatcode{make\_restriction\_operator} and \formatcode{make\_prolongation\_operator} in the \formatcode{bspline.gridops} module, and precomputing $J$ is done by \formatcode{compute\_j\_zeroplus} in the \formatcode{bspline.coefficients} module.
The operator implementations exploit the separable structure of $J_{\gridpointIndexN} = J_{n_x} J_{n_y} J_{n_z}$ to perform them one coordinate direction at a time.

Third, the computation of the kernel coefficient stencils $\kernelStencil^{\levelIndex}$, unspecified until now, is another key part of any particular choice of interpolation scheme.
It is known that the coefficients of the B-spline interpolant can be obtained as the convolution of values of the target function on grid points with a universal sequence.\cite{Sch73}
For interaction kernels of the considered type, in three dimensions, the calculation of the coefficient stencils reads
\begin{equation}\label{eq:kernel_stencil_by_conv_with_universal_coeff}
  \kernelStencil^{\levelIndex}_{\gridpointIndexM} = \sum_{n_x}\omega'_{m_x-n_x} \sum_{n_y}\omega'_{m_y-n_y} \sum_{n_z}\omega'_{m_z-n_z} \, k_l(\mathbf{h}_{\levelIndex} \cdot \gridpointIndexN)
  \, ,
\end{equation}
with $\mathbf{h}_{\levelIndex} = (h_{\levelIndex,x}, h_{\levelIndex,y}, h_{\levelIndex,z})\transpose$, i.e., the convolution can be done one coordinate direction at a time.
A key contribution of Ref.~\citenum{HWX+16} was the development of an algorithm that brings down the cost of precomputing the universal coefficients $\omega'_n$ by dropping the requirement of exactly matching the target function at grid points, while preserving control over the overall approximation accuracy, resulting in \emph{quasi-interpolation}.
We implemented this algorithm in \programnametext, making heavy use of symbolic computation by way of the \formatcode{sympy} Python package.\cite{sympy2017}
This being a preprocessing step, there is no need for it to be compatible with \formatprogramtitle{JAX} transformations.
The implementation is in the \formatcode{bspline.coefficients} module.

\subsection{Splitting for Coulomb kernels}
\label{ssec_coulomb_kernel_splitting}
In \programnametext, we implement the same splitting of the Coulomb kernel as in Ref.~\citenum{HWX+16}.
The hitherto unspecified partial kernels in \cref{eq:kernel_splitting} now take the form
\begin{equation}\label{eq:coulomb_splitting_with_softener}
  k_\levelIndex(r) =
    \begin{cases}
      \frac{1}{r} - \frac{1}{\levelZeroCutoff} \gamma\left(\frac{r}{\levelZeroCutoff}\right)   & \levelIndex = 0 , \\
      \frac{1}{2^{\levelIndex - 1} \levelZeroCutoff} \gamma\left(\frac{r}{2^{\levelIndex - 1} \levelZeroCutoff}\right) - \frac{1}{2^{\levelIndex} \levelZeroCutoff} \gamma\left(\frac{r}{2^{\levelIndex} \levelZeroCutoff}\right)   & 1 \leq \levelIndex \leq L - 1 \, , \\
      \frac{1}{2^{L - 1} \levelZeroCutoff} \gamma\left(\frac{r}{2^{L - 1} \levelZeroCutoff}\right)            & \levelIndex = L \, ,
    \end{cases}
\end{equation}
with a dimensionless so-called \emph{softening function} $\gamma(\rho)$ that is required to be $\gamma(\rho) = 1 / \rho$ for $\rho \geq 1$, and bounded and smooth for $\rho \leq 1$.
It is shown in Ref.~\citenum{HWX+16} that, for a given interpolation order $p$, an in some sense optimal choice for the part of $\gamma(\rho)$ with $\rho \leq 1$ is obtained by Taylor-expanding $s^{-1/2}$ around $s=1$ up to polynomial degree $p - 1$, followed by setting $s = \rho^2$.

The relevant components of \programnametext, all in the \formatcode{\programnamecode.kernels} module, are the following:
\begin{enumerate}
  \item \formatcode{split\_one\_over\_r}: This creates a list of functions, one for each level and representing one of the partial kernels, according to the logic of \cref{eq:coulomb_splitting_with_softener}.
  \item \formatcode{SoftenerOneOverR}: This is a callable class that implements the concrete softening function $\gamma(\rho)$.
  \item \formatcode{make\_construct\_stencils}: This sets up a \formatcode{kernel\_stencil\_construction\_fn} appropriate for passing to \formatcode{make\_compute\_u\_oneplus} (see \cref{sssec_impl_core_longrange} and the bottom panel of \cref{fig:core_modularity}).
    The returned function computes the kernel stencils $\kernelStencil^{\levelIndex}$ according to \cref{eq:kernel_stencil_by_conv_with_universal_coeff} and exploits the fact that the intermediate partial kernels ($\levelIndex = 1, \dots, L - 1$) in \cref{eq:coulomb_splitting_with_softener} all have the same functional form, differing only in the prefactor and scaling of the argument.
    The $\kernelStencil^{\levelIndex}$ for $\levelIndex = 2, \dots, L-1$ are therefore computed by simply repeatedly dividing $\kernelStencil^1$ by two.
\end{enumerate}
Extensions to a different interaction kernel or different splitting of the same kernel would need to replace (at least) these three components with custom implementations, but they would remain a useful starting point.

\subsection{High-level interface}
\label{ssec_high_level_interface}
The \formatcode{\programnamecode.calculators} module provides a convenience wrapper to use the MSM implementation with B-spline interpolation for Coulomb kernels from \cref{ssec_bspline_interp,ssec_coulomb_kernel_splitting}.
It is built around a Python data class \formatcode{MSMParams} as its central data structure for holding all required settings.
For convenient debugging, human-readable settings documentation, and restarting or reproducing calculations, \formatcode{MSMParams} has two methods \formatcode{save\_json} and \formatcode{load\_json} for JSON (de-)serialization.
Passing an \formatcode{MSMParams} instance to the function \formatcode{create\_msm} returns a dictionary of evaluation functions for various quantities:
\begin{lstlisting}[language=Python,firstline=7]
msm_params = MSMParams.load_json("params.json")
evaluation_fns = create_msm(msm_params)
eval_energy_and_forces = jax.jit(
    evaluation_fns["energy_and_forces"]
)
energy, forces = eval_energy_and_forces(
    positions, charges
)
\end{lstlisting}

While manual instantiation of \formatcode{MSMParams} is possible, the simplified interface for doing so provided by the helper function \formatcode{set\_up\_msm\_params} should be preferred in most cases.
It has seven required arguments, three of which, \formatcode{level\_zero\_cutoff}, \formatcode{level\_one\_spacings}, and \formatcode{p}, constitute the main algorithm parameters.
The remaining four are geometry-related.
They are \formatcode{cell}, \formatcode{pbc} (periodicity along each axis), \formatcode{cell\_mode} (specifying simplifying assumptions on the shape of the unit cell), and \formatcode{dynamic\_cell} (whether the cell can change).

\subsection{Units}
\label{ssec_units}

\programnametext{} does not assume any specific physical units.
It computes the numerical value of the expression \cref{eq:generic_pair_potential_sum}, and its derivatives, from whatever numerical values it received as the inputs for positions and charges.
It is the user's responsibility to convert the outputs to the desired units.

\section{Examples}
\label{sec_examples}

\subsection{Madelung constants}
\label{ssec_madelung}

A natural first sanity check on any algorithm or code for the calculation of electrostatics in periodic systems is the evaluation of Madelung constants.\cite{Mad18}
They provide reference values for electrostatic lattice energies of simple and well-defined test systems and are tabulated to high accuracy.

Several slightly different definitions of the Madelung constant are in use,\cite{Qua70} which must be kept in mind when comparing calculated values with tabulated ones.
The one used in this example is defined by
\begin{equation}\label{eq:madelung_energy}
  U = - N_\text{f.u.} \times M \times \frac{z^2 e^2}{d_{\text{min}}} \, ,
\end{equation}
where $U$ is the electrostatic lattice energy per unit cell, $N_\text{f.u.}$ is the number of formula units contained therein, $e$ is a unit of charge, and $d_{\text{min}}$ is the minimum cation-anion distance for the crystal structure under consideration.
Writing the cation and anion charges as $z_\text{cation} e$ and $- z_\text{anion} e$, respectively, and defining $z = \operatorname{gcd}(z_\text{cation}, z_\text{anion})$, the meaning of $(z^2 e^2) / (d_{\text{min}})$ is that of a characteristic reference energy for a \emph{single} ion pair.
The remaining dimensionless proportionality factor $M$ relating it to the lattice energy is the Madelung constant.

We calculated Madelung constants for six structures: NaCl in the conventional unit cell, CsCl, zincblende, CaF$_2$, NaCl in the primitive unit cell, and wurtzite (with $c/a = \sqrt{8/3}$ and $u=3/8$).
The last two demonstrate the ability of \programnametext{} to handle non-orthorhombic unit cells.
Reference values for the Madelung constants were taken from Refs.~\citenum{Hou20} (wurtzite) and \citenum{Gla12} (all other structures).

\begin{figure*}
  \centering
  \includegraphics[width=\textwidth]{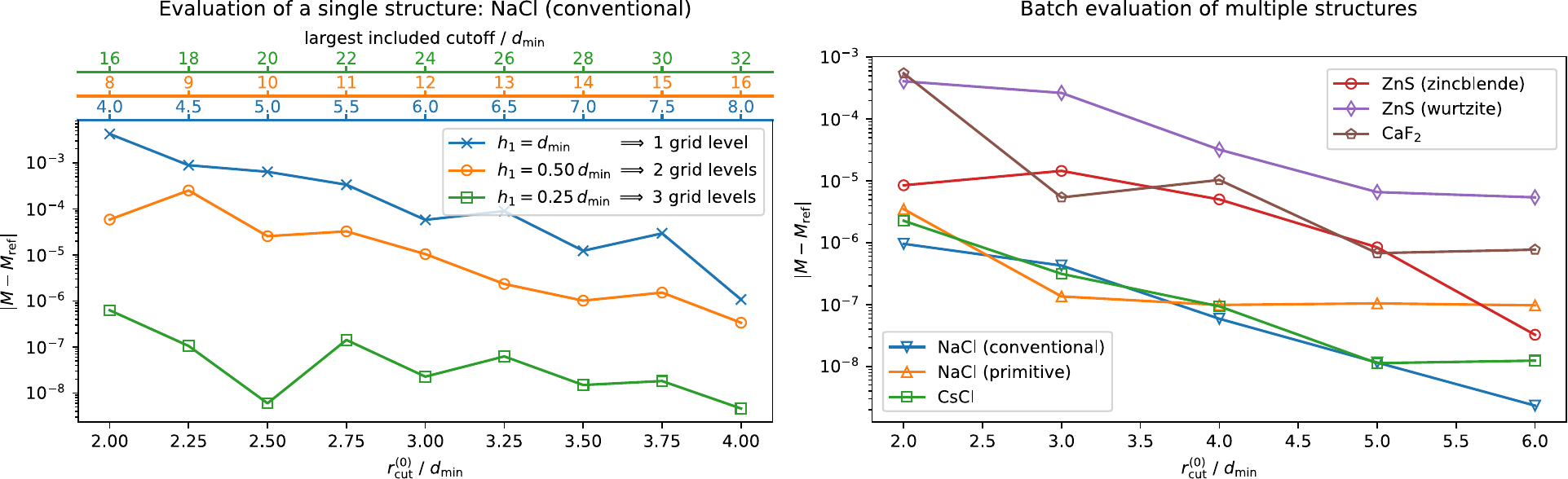}
  \caption{
    \label{fig:madelung_err_vs_cutoff}
    Absolute deviations between calculated Madelung constants $M$ and reference values $M_{\mathrm{ref}}$, plotted as a function of the MSM's level-zero cutoff relative to each structure's respective minimum cation-anion distance.
    All results shown were calculated using interpolation order $p=6$ and double-precision arithmetic.
    Left: Results for NaCl (conventional cell), calculated using models set up for this one structure specifically.
    Different lines correspond to different grid spacings and accordingly different numbers of levels.
    Right: Results for all six included structures, calculated using models set up so as to be mappable over all structures.
    Different lines correspond to different structures.
  }
\end{figure*}

Two different example scripts for calculating Madelung constants can be found in \formatcode{examples/madelung\_constants/} and are called \formatcode{calculate\_madelung\_single.py} and \formatcode{calculate\_madelung\_batch.py}.
The first evaluates the Madelung constants for one structure at a time, for each structure varying the MSM's level-zero cutoff radius $r_{\mathrm{cut}}^{(0)}$ as well as the level-one grid spacing $h_1$.
Results of this are provided for all structures in the \formatcode{example\_results/single/} subdirectory, and one representative figure (for conventional-cell NaCl) is shown in the upper panel of \cref{fig:madelung_err_vs_cutoff}.
With an interpolation order $p=6$, we attain accuracies of seven to eight decimals for all structures, except for wurtzite, where the number of decimals of the reference value is only five to begin with.

On top of serving as a correctness check, we also intend for this first example to illuminate the grid setup process in periodic systems, in which case the grid spacing keeps being doubled until there is only a single point at the highest level.
Halving the grid spacing therefore has the effect of increasing the total number $L$ of levels by one.
Since the cutoff doubles alongside the spacing from level to level, for a given level-zero cutoff $\levelZeroCutoff$, the largest effective cutoff radius $r_{\mathrm{cut}}^{(L-1)} = 2^{L-1} \levelZeroCutoff$ included in the evaluation is doubled as well.
Making corresponding changes to $\levelZeroCutoff$ and $h_1$ thus amounts to shifting computational effort between the directly-evaluated and grid parts, not unlike how the Gaussian width parameter in Ewald(-based) methods controls the tradeoff between direct- and reciprocal-space evaluation.

\begin{figure}[htb]
  \centering
  \includegraphics[width=\columnwidth]{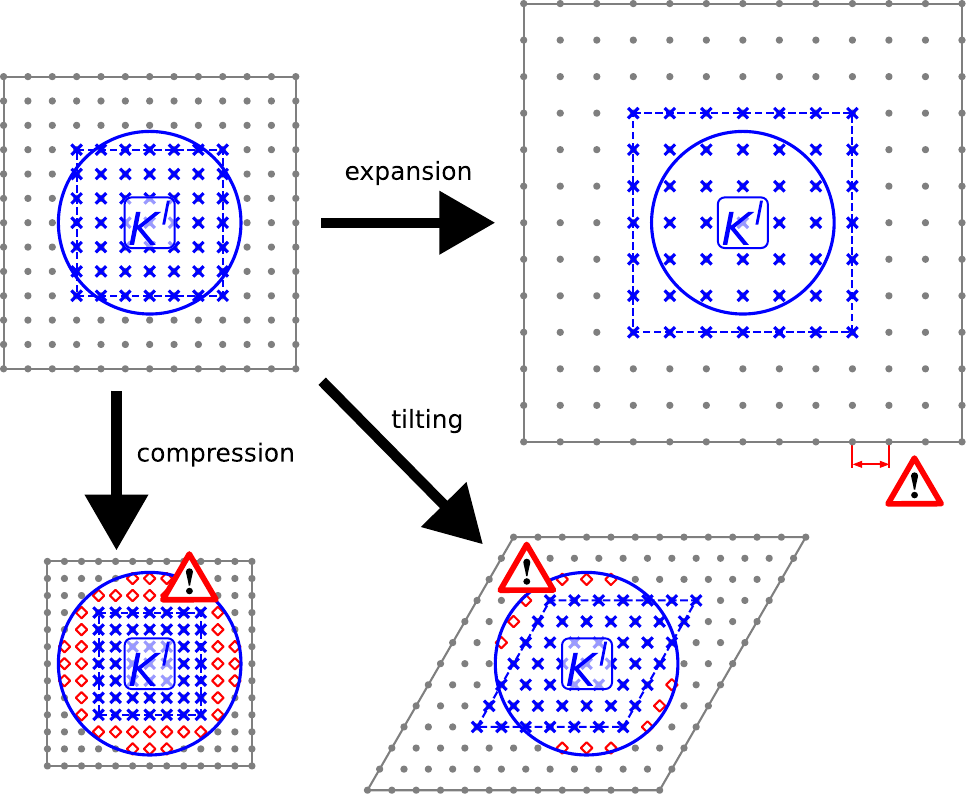}
  \caption{
    \label{fig:stencils_cell_shape_changes}
    Schematic of the different failure modes causing a loss of accuracy in the grid part of the calculation when evaluating structures having different cells.
    Expansion decreases the resolution of a fixed-size grid as the grid spacing scales alongside the cell.
    Compression or tilting can cause a fixed-size kernel stencil $\kernelStencil^{\levelIndex}$ (array of blue crosses with dashed exterior outline) to no longer capture the relevant cutoff radius of the potential (highlighted by empty red diamonds).
  }
\end{figure}

The crystal structures considered in this example are disparate both in the numbers of ions contained and in the geometries of their unit cells, necessitating different values for model parameters like number of grid points, number of levels, and size of the stencil arrays $\kernelStencil^{\levelIndex}$.
Normally, due to \formatprogramtitle{JAX} imposing the requirement of static array shapes inside JIT-compiled functions, it would be necessary to set up and compile a separate evaluation function for each crystal structure.
However, it is possible to work around this issue by applying some preprocessing to the input arrays and choosing the model parameters with care.
This is demonstrated by the second example script, \formatcode{calculate\_madelung\_batch.py}, which uses \formatprogramtitle{JAX}'s \formatcode{vmap} transform to vectorize a \emph{single} appropriately defined \programnametext{} model over all structures.
Results are shown in the bottom panel of \cref{fig:madelung_err_vs_cutoff}.
To make the model work across structures, the main challenges to be considered are the following:
\begin{enumerate}
  \item Handling different numbers of atoms between structures:
    Perhaps the most obviously problematic issue, as it obviously violates the static-array-shape requirement, this is solved by padding all input structures to the same number of atoms with placeholder atoms that have charges of zero and therefore do not change the electrostatic energy.
  \item Ensuring appropriate grid spacing:
    Due to the requirement of static array shapes, what is ultimately fixed in an \programnametext{} model is not the grid spacing but the number of grid points along each direction.
    The resulting actual spacings scale with the input unit cell, so when using the same model for different structures one needs to make sure it leads to reasonable (not too large) spacings for each of them.
    The problem is illustrated by the \enquote{expansion} scenario in \cref{fig:stencils_cell_shape_changes}.
    In this example, we aimed for spacings satisfying $h_1 \lesssim d_{\mathrm{min}}$ in all structures.
  \item Accommodating the cutoff during the evaluation of the short-range (direct) part:
    In the evaluation of \Uzero, the minimum-image convention is used.
    For this to result in correct pair distances, the level-zero cutoff must satisfy $\levelZeroCutoff \leq V / (2 A_{\mathrm{max}})$, where $V$ is the volume of the cell and $A_{\mathrm{max}}$ the maximum area among its faces.
    If a given cell would otherwise be too small, this is remedied by evaluating the interaction in a supercell.
    The number of replicas of the original cell comprised in that supercell must be known at compile time (again because of static-shape requirements) and must be set to a large enough value to work across all structures to be evaluated.
  \item Accommodating the cutoff during the evaluation of the long-range (grid) part:
    This is in some sense the analog of the third point for the grid part of the calculation, and concerns the size of the kernel stencil arrays $\kernelStencil^{\levelIndex}$ from \cref{eq:U_oneplus_grid_approx}.
    The different structures in this example having different actual grid spacings (as per the second point) as well as cell vector tilts leads to different numbers of points being required to cover the potential cutoffs.
    The problem is illustrated by the scenarios labeled \enquote{compression} and \enquote{tilting} in \cref{fig:stencils_cell_shape_changes}.
    To work for all structures, the common stencil size must be set to the largest out of all structures.
\end{enumerate}
Even though such an approach means using larger input structures (i.e., containing more particles) and more expensive settings than would be necessary for some of the structures on their own, avoiding expensive function recompilations and gaining the ability to vectorize can greatly outweigh the extra cost, depending on the application.

\subsection{Benchmarks}
\label{ssec_benchmarks}
In this section we briefly present the results of a number of performance benchmarks, contained in \formatcode{examples/benchmarks}.

These benchmarks were performed on synthetic data comprised of particles placed randomly inside cubic boxes with a constraint on the minimum distance between two particles.
This avoids high forces distorting accuracy metrics, as well as needing unnecessarily and unrealistically large neighbor lists.
To satisfy the distance constraint, a two-stage data generation procedure was used.
First, initial configurations were generated using \formatprogramtitle{PACKMOL},\cite{packmol2009} called through a convenience wrapper taken from the \formatprogramtitle{Clinamen2} package.\cite{WBK+24}
Second, since \formatprogramtitle{PACKMOL} does not take into account periodic boundary conditions, we performed a post-processing step in which a short-range repulsive pair potential was put on the particles, and particles near the boundary were relaxed under periodic boundary conditions, with the rest kept fixed.
If we define $\overline{d} = (V / N)^{1/3}$, where $V$ is the volume and the exponent is related to the spatial dimension, the end result of the procedure is that no two particles are closer than $0.6 \overline{d}$.
Uniform random charges between $-1$ and $+1$ (arbitrary units) were placed on the particles, followed by subtracting the average charge from each one to achieve overall charge neutrality.
The final configurations, which are also used in some of the unit tests, are located under \formatcode{\programnamecode/data/}, alongside the code used to generate them.

All benchmarks were run on an NVIDIA A40 GPU, using \formatcode{jax} and \formatcode{jaxlib} version 0.7, and version 12.9 of the CUDA toolkit.
Evaluation times were measured using Python's built-in \formatcode{timeit.repeat} function, with the \formatcode{repeat} and \formatcode{number} parameters chosen individually for each kind of benchmark such that it finishes running in a reasonable amount of time.
Times do not include the time taken to construct the neighbor list for the short-range part, nor do they include the time to copy data to or from the GPU.

\subsubsection{Accuracy-performance tradeoff}
\label{sssec_accuracy_vs_performance}
This example, located in the \formatcode{cost\_vs\_accuracy/} subdirectory of \formatcode{benchmarks/}, analyzes the relationship between the accuracy achieved by \programnametext{} and the associated computational cost as a function of the method parameters.
Ref.~\citenum{HWX+16}, on which our implementation is based, contains a similar analysis.
Given the differences in implementation and performance characteristics of GPUs compared to CPUs, we deemed it in order to revisit it.

We performed the benchmark for two distinct kinds of systems, a medium-sized one ($10^4$ particles) with no periodicity, and a small one ($10^2$ particles) with periodic boundary conditions.
For both kinds, we calculated various target quantities using \programnametext{} with different settings, for 11 different particle configurations, and recorded the average evaluation time alongside the accuracy compared to results obtained with a reference method.
In the non-periodic case, where this is indeed possible, the reference results were calculated exactly, whereas in the periodic case we obtained them using the P3M method as implemented in \formatprogramtitle{LAMMPS}, with its accuracy parameter set to $10^{-8}$ (``\formatcode{kspace\_style pppm 1e-08}'' command).

\begin{figure}[htb]
  \centering
  \includegraphics[width=\columnwidth]{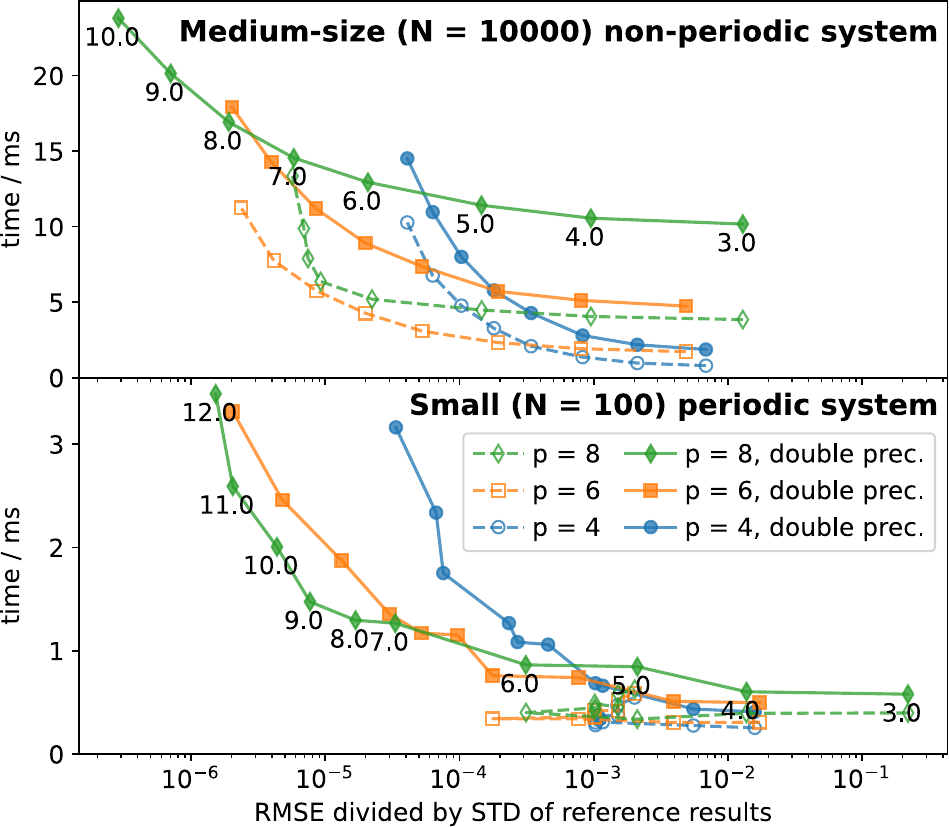}
  \caption{
    \label{fig:cost_vs_accuracy}
    Each point shows the error in the forces calculated with \programnametext{} (x-axis) and the corresponding evaluation time (y-axis) for one combination of method parameters.
    The error metric used is the root mean square error (RMSE) between the results from \programnametext{} and from a reference method (see main text), further divided by the standard deviation (STD) of the reference results to put it in relation to the spread of values in the data.
    Within the individual curves, the level-zero cutoff $\levelZeroCutoff$ is varied for otherwise identical settings, its values in units of the average neighbor particle spacing $\overline{d}$ indicated by the text annotations of the points.
  }
\end{figure}

The results for forces are shown in \cref{fig:cost_vs_accuracy} for a range of interpolation orders $p$, level-zero cutoff radii $\levelZeroCutoff$, as well as for single- and double-precision arithmetic.
To keep the parameter space easily visualizable, the level-one grid spacing $h_1$ was set to the average interparticle spacing $\overline{d}$ throughout, in line with the recommendation from Ref.~\citenum{HWX+16}, but it would not be difficult to extend the benchmark to scan over it as well.
Besides the results shown for forces, the example also supports, and includes sample results of, the evaluation of energies, charge gradients, and stress tensors.

Conclusions that can be drawn from this test include that using interpolation order $p=8$ is beneficial only in conjunction with double-precision arithmetic, and that when $p=4$ it is not advantageous to increase the cutoff beyond around six to seven times $\overline{d}$, at which point rather $p$ itself should be increased.
The equivalent crossover when $p=6$ occurs at cutoffs around nine to ten times $\overline{d}$.
An exception to these trends is seen for the small periodic system in single precision, where, up to around six times $\overline{d}$, increasing the cutoff improves the accuracy while the run time stays constant, indicating that it is dominated by a constant overhead.
Beyond that, further increasing the cutoff does not improve the accuracy.
The above conclusions notwithstanding, we generally recommend that users pick algorithm parameters by running benchmarks on their own hardware specifically for the systems and quantities that are of interest to them.
The provided benchmark scripts are designed as an easily adaptable starting point for doing so.

\subsubsection{Scaling with system size}
\label{sssec_scaling}

The MSM's $\mathcal{O}(N)$ scaling is a theoretically rigorous result.
However, this holds only strictly in the asymptotic limit, and in practice any implementation on real hardware incurs overheads and non-idealities.
This example, in the \formatcode{scaling/} subdirectory of \formatcode{examples/benchmarks/}, therefore investigates the real-world scaling of \programnametext{} with the number of particles.

\begin{figure}[htb]
  \centering
  \includegraphics[width=\columnwidth]{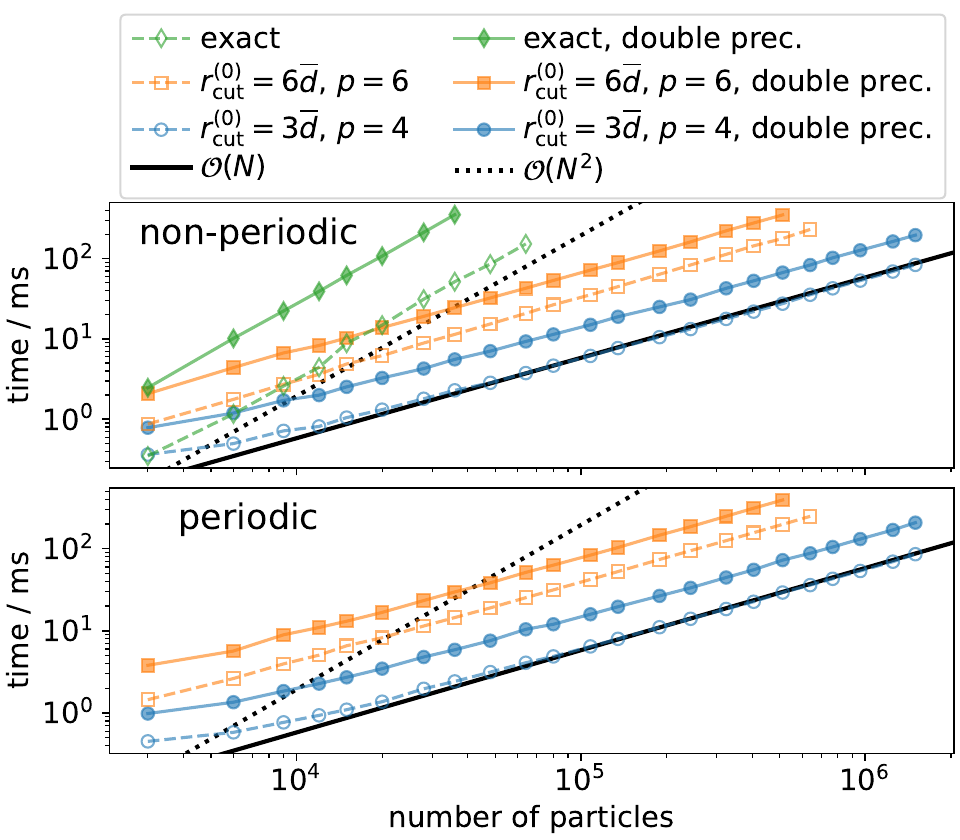}
  \caption{
    \label{fig:scaling_benchmark}
    Timings of force evaluation as a function of the number of particles, for different MSM settings, and in both single and double precision.
    For the case with no periodicity, the timings of the exact calculation via explicit evaluation of all pairs are also included for comparison (green lines in the upper panel).
    Where the graphs end at a lower number of particles than the overall highest one included, this indicates that the calculation ran out of memory.
    Additionally included for reference are examples of perfect linear and quadratic scaling (solid and dotted black lines).
  }
\end{figure}

To that end, we recorded the times to evaluate electrostatic forces using two different sets of MSM parameters, both with and without periodicity, in systems ranging in size from $N = 3000$ to $N = 1.5 \times 10^6$.
While structures with $N \leq 15000$ were generated as described at the beginning of \cref{ssec_benchmarks}, all larger structures in this test were obtained by making differently sized supercells from the former.

For the non-periodic case, the only case where this is feasible, the graph in \cref{fig:scaling_benchmark} also includes the timings of the exact $\mathcal{O}(N^2)$-scaling calculation, i.e., the explicit summation over all pairs.
For the sake of both convenience and fair comparison, the exact calculation was performed with \formatcode{\programnamecode{}.core.shortrange.make\_eval\_pair\_pot}, which is itself part of \programnametext{}.
The contrast between this and the MSM clearly shows the latter's superior scaling, that is consistent with asymptotic $\mathcal{O}(N)$ behavior.
It is worth noting not just the differences in run time, but also memory usage, with the MSM able to handle an order of magnitude more particles before running out of memory.

\subsubsection{Custom derivatives of long-range part}
\label{sssec_benchmarks_custom_deriv}

\begin{figure}[htb]
  \centering
  \includegraphics[width=\columnwidth]{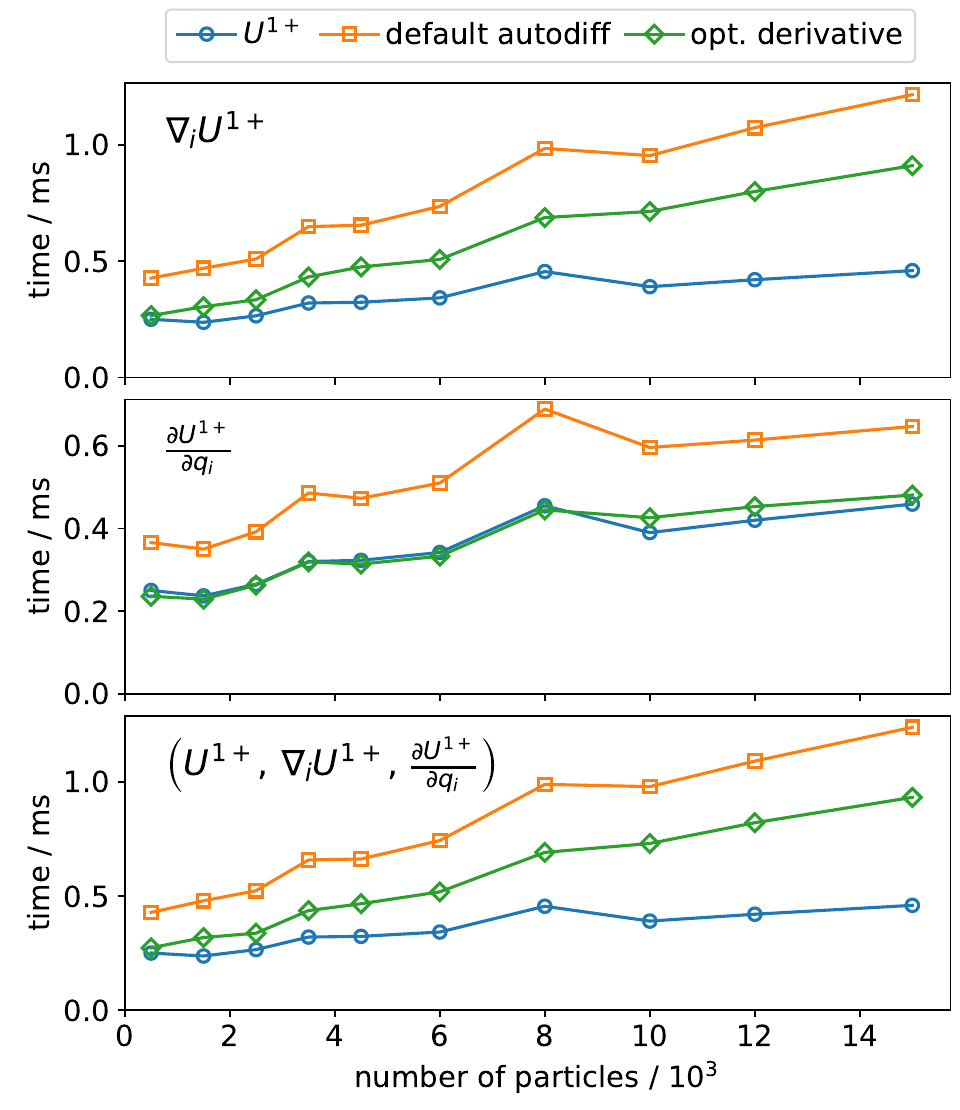}
  \caption{
    \label{fig:timing_default_vs_custom_grad}
    Comparison of evaluation times of the derivatives of the long-range energy \Uoneplus{} using standard automatic differentiation (orange squares) and optimized derivative rules (green diamonds).
    The top and middle panels contain results for the gradients w.r.t.~positions and charges (with \formatcode{jax.grad}), respectively.
    The bottom panel shows the simultaneous evaluation of the energy and both its position and charge gradients (with \formatcode{jax.value\_and\_grad}).
    All panels also include the times for evaluating the energy alone (blue circles, these data are the same in all panels), to give a sense of the extra cost of calculating derivatives over just the energy.
  }
\end{figure}

This example, in the \formatcode{custom\_longrange\_derivative/} subdirectory of \formatcode{examples/benchmarks/}, demonstrates the practical gains due to the custom derivative rules for \Uoneplus{} described in \cref{sssec_impl_core_longrange}.
The results shown in \cref{fig:timing_default_vs_custom_grad} were obtained for periodic systems in single-precision arithmetic, with $\levelZeroCutoff = 3 \overline{d}$, $h_1 \approx \overline{d}$, $p = 4$.
Note that there is no neighbor list involved, because only the grid part of the algorithm is evaluated.

The results suggest that, for the gradient w.r.t.~positions, the difference between default automatic differentiation and the optimized implementation based on \cref{eq:analytic_uoneplus_position_grad} is a more or less constant offset, and the advantage of the custom implementation is thus relatively more pronounced at low to medium particle numbers.
The cost of calculating the gradient w.r.t.~charges, on the other hand, becomes the same as evaluating the energy itself when the custom derivative rule based on \cref{eq:analytic_uoneplus_charge_grad} is used.
This matches the expected behavior as the value of the charge gradient appears in the energy expression \cref{eq:uoneplus_interpolation_with_index_set} as an intermediate result that dominates its cost.

\subsection{MD simulation of charged Lennard-Jonesium}
\label{ssec_charged_lj_md}

This example, in \formatcode{examples/charged\_lj\_md/}, demonstrates the use of \programnametext{} in MD simulations, and acts as another check of correctness, as some subtle failures may conceivably not have shown up during the previous sections' controlled experiments on purely static systems.
Moreover, it provides a template for integration of \programnametext{} with the atomic simulation environment (\formatprogramtitle{ase}) package,\cite{HJB+17} which we used to perform the simulations.

We picked a simple model system consisting of $N = 1000$ particles in a periodic cubic box that have fixed charges and interact over short distances via an additional Lennard-Jones (LJ) potential.
We do not aim for a quantitative study in this example, only to demonstrate the stability of MD simulations with \programnametext{}.

Charged LJ systems have been studied in the past as idealized models for ionic fluids.\cite{AK17}
However, here 
we use it as a generic placeholder for the combination of a short-range potential, that could be replaced, e.g., by a short-range MLIP, and a long-range potential. 
The total system energy reads
\begin{equation}\label{eq:charged_lj_energy}
  \begin{gathered}
    U = \frac{1}{2} \sum_{i=1}^N \sum_{j \neq i}
      \left[
        u_{\mathrm{LJ}}(r_{ij}) \times f_{\mathrm{c}}(r_{ij}) + u_{\mathrm{coul}}(r_{ij}, q_i, q_j)
      \right] \, ,
    \\
    u_{\mathrm{LJ}}(r_{ij}) = 4 \varepsilon
      \left[
        \left(\frac{\sigma}{r_{ij}}\right)^{12} - \left(\frac{\sigma}{r_{ij}}\right)^6
      \right]
      \, ,
    \\
    u_{\mathrm{coul}}(r_{ij}, q_i, q_j) = \frac{q_i q_j}{r_{ij}} \, ,
  \end{gathered}
\end{equation}
where $\varepsilon$ and $\sigma$ are the LJ potential's characteristic energy and length scales, respectively, and $f_{\mathrm{c}}$ is a cutoff function that goes to zero smoothly.
We used the same cutoff radius value $r_{\mathrm{cut}} = 4 \sigma$ in both $f_{\mathrm{c}}$ and the short-range part of the MSM.
The charges $q_i$ differ between particles only in sign (positive for one half, negative for the other), and their magnitudes were set such that $\left| u_{\mathrm{coul}}(r_{ij}=\sigma, q_i, q_j) \right| / \varepsilon = 50$.

Similar to the structures used for benchmarking throughout \cref{ssec_benchmarks}, we used \formatprogramtitle{PACKMOL} for initial configuration setup.
This was followed by an energy minimization, and initialization of velocities with a distribution corresponding to twice the target temperature.
Starting from this, we ran NVE and NPT simulations.
In the latter case, this was preceded by a warm-up NVT run, the averaged pressure from the second half of which was then set as the target pressure for the NPT run.
The integrator classes from \formatprogramtitle{ase} used for the NVE, NVT, and NPT simulations are \formatcode{VelocityVerlet}, \formatcode{NoseHooverChainNVT}, and \formatcode{IsotropicMTKNPT}, respectively, with a time step of \SI{1}{\femto\second} and otherwise default settings.
We decided not to set up and run the simulations in reduced LJ units (where quantities are made dimensionless by scaling with $\varepsilon$, $\sigma$, or derived quantities), but instead performed them for an artificial system of LJ-parametrized argon atoms ($\epsilon = \SI{125.2}{\kelvin} / k_{\mathrm{B}}$ and $\sigma = \SI{3.405}{\angstrom}$, from Ref.~\citenum{Whi99}) to which we added charges.
The reason is that we found this a more convenient way to consistently supply inputs in the units expected by \formatprogramtitle{ase}.

\begin{figure}[htb]
  \centering
  \includegraphics[width=\columnwidth]{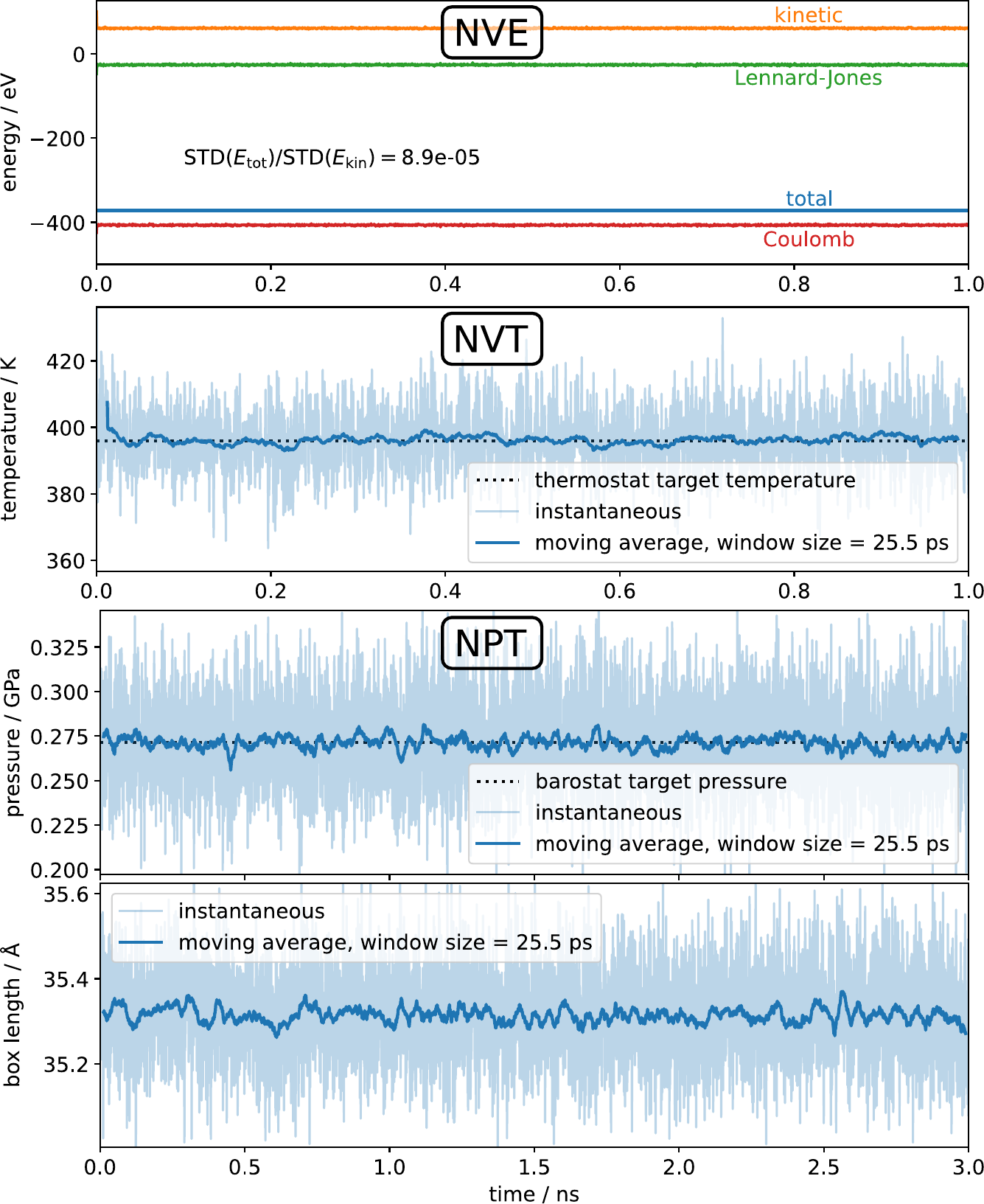}
  \caption{
    \label{fig:charged_lj_md_different_ensembles}
    Various quantities along trajectories in MD simulations of charged Lennard-Jones particles.
    The numerical values in physical units are the result of an ultimately arbitrary and inconsequential choice for the Lennard-Jones interaction parameters and only meaningful in relation to these parameters, see main text for explanation.
    The NVE plot (top panel) includes the ratio between the standard deviations (STD) over the trajectory of total and kinetic energy as a measure of stability.
  }
\end{figure}

Results from runs in the different ensembles are shown in \cref{fig:charged_lj_md_different_ensembles}.
The initial particle density in these runs was $n = 0.85 / \sigma^3$, and the target temperature $T = 3.15 \, \varepsilon / k_{\mathrm{B}}$.
No significant energy drift was observed in the NVE simulation, which we quantified via the ratio between the standard deviations of total and kinetic energy, and temperature, pressure, and the box size (in NPT) can be seen to have remained stable as well.

Note that, when setting up \programnametext{} for use in NPT simulations similar considerations arise as described already for the batch-mode evaluation in \cref{ssec_madelung} and illustrated in \cref{fig:stencils_cell_shape_changes}.
It is good practice to monitor the cell shape and size over the course of an MD trajectory and verify that it remains within the safe limits supported by the chosen MSM parameters.

\section{Conclusions}
\label{sec_conclusions}
We have presented \programnametext, a Python package that provides an implementation of the multilevel summation method (MSM) built on the \formatprogramtitle{JAX} framework, with a view toward being a suitable building block for machine-learned interatomic potentials (MLIPs) in materials science.
We have discussed the package structure and implementation choices in some detail, in a way that we hope will enable users who want to adapt or extend the code to get started quickly.
Through various examples, we have verified that \programnametext{} accurately calculates electrostatic energies and derived quantities in periodic and nonperiodic systems.
Based on benchmarks for specific systems we have made tentative recommendations on optimal method parameter choices, and pointed out utilities for users to run their own parameter tuning, if needed.
Moreover, we have verified that, in practice, our implementation indeed achieves the asymptotic linear scaling with the number of particles that the MSM is theoretically predicted to have.
Lastly, we have demonstrated the stability of \programnametext{} when used as (part of) a force field in MD simulations.

\section*{Declaration of competing interest}
There are no competing interests to declare.

\section*{Data availability}
The latest version of \programnametext{} can be installed from PyPi via ``\formatcode{pip install msmjax}'', or cloned from \href{https://github.com/Madsen-s-research-group/msmjax}{https://github.com/Madsen-s-research-group/msmjax}.
The version described and used in the present article is tagged in that repository as \href{https://github.com/Madsen-s-research-group/msmjax/releases/tag/submission_cpc_20251006}{\formatcode{submission\_cpc\_20251006}} (also registered on Zenodo under \href{https://doi.org/10.5281/zenodo.17279444}{https://doi.org/10.5281/zenodo.17279444}).
The repository version contains all scripts and input files required for running the examples described in this article.
It also contains sample outputs for all of them, with the exception of the MD trajectories from the charged-Lennard-Jones example in \cref{ssec_charged_lj_md}, which, for file size reasons, are not included with the code, but are available under \href{https://doi.org/10.5281/zenodo.17279790}{https://doi.org/10.5281/zenodo.17279790}.
The documentation is under \href{https://madsen-s-research-group.github.io/msmjax/}{https://madsen-s-research-group.github.io/msmjax/}.

\section*{Acknowledgements}
This research was funded in part by the Austrian Science Fund (FWF) 10.55776/F8100 and 10.55776/COE5. For open access purposes, the author has applied a CC BY public copyright license to any author accepted manuscript version arising from this submission. It was also supported by MCIN with funding from the European Union NextGenerationEU (PRTR-C17.I1) promoted by the Government of Aragon. J.C. acknowledges Grant CEX2023-001286-S funded by MICIU/AEI /10.13039/501100011033.




\biboptions{sort&compress}
\biboptions{super}
\bibliographystyle{elsarticle-num}
\bibliography{msm_bibliography_zotero.bib,msm_bibliography_external.bib}







\end{document}

%% file: preamble.tex
\documentclass[final,3p,times,twocolumn]{elsarticle}

\makeatletter
\def\ps@pprintTitle{%
 \let\@oddhead\@empty
 \let\@evenhead\@empty
 \def\@oddfoot{}%
 \let\@evenfoot\@oddfoot}
\makeatother

\usepackage[version=4]{mhchem}

\usepackage{amssymb}
\usepackage{amsmath}
\usepackage[noend]{algpseudocode}
\usepackage{algorithm}

\usepackage[usenames,dvipsnames]{xcolor}
\definecolor{jazzberryjam}{rgb}{0.65, 0.04, 0.37}

\usepackage[hidelinks]{hyperref}

\usepackage[capitalize,nameinlink]{cleveref}

\definecolor{ivory}{RGB}{255,255,235}
\definecolor{lighter-gray}{gray}{.85}
\usepackage{listings}
\definecolor{codegreen}{rgb}{0,0.6,0}
\definecolor{codegray}{rgb}{0.5,0.5,0.5}
\definecolor{codepurple}{rgb}{0.58,0,0.82}
\definecolor{backcolour}{rgb}{0.95,0.95,0.92}
\lstdefinestyle{mystyle}{
    backgroundcolor=\color{backcolour},   
    commentstyle=\color{codegreen},
    keywordstyle=\color{magenta},
    numberstyle=\tiny\color{codegray},
    stringstyle=\color{codepurple},
    basicstyle=\ttfamily\footnotesize,
    breakatwhitespace=false,         
    breaklines=true,                 
    captionpos=b,                    
    keepspaces=true,                 
    numbers=left,                    
    numbersep=5pt,                  
    showspaces=false,                
    showstringspaces=false,
    showtabs=false,                  
    tabsize=2
}

\usepackage{siunitx}
\DeclareSIUnit\angstrom{\text {Å}}
\DeclareSIUnit\atom{atom}

\newcommand{\formatprogramtitle}[1]{\textsc{\mbox{#1}}}
\newcommand{\formatcode}[1]{\texttt{\mbox{#1}}}

\usepackage{csquotes}

\newcounter{bla}

\newcommand{\programnametext}{\formatprogramtitle{msmJAX}}
\newcommand{\programnamecode}{msmjax}

\newcommand{\levelZeroCutoff}{\ensuremath{r_{\mathrm{cut}}^{(0)}}}

\newcommand{\Uoneplus}{\ensuremath{U^{1+}}}
\newcommand{\Uzero}{\ensuremath{U^0}}

\newcommand{\posvec}{\mathbf{r}}

\newcommand{\gridCharge}{\ensuremath{\widetilde{q}}}
\newcommand{\operatorK}{\ensuremath{\mathcal{K}}}
\newcommand{\kernelStencil}{\ensuremath{K}}
\newcommand{\operatorI}{\ensuremath{\mathcal{I}}}
\newcommand{\transpose}{^{\mathrm{T}}}
\newcommand{\sumParticlesI}{\sum_{i=1}^N}
\newcommand{\sumParticlesJ}{\sum_{j=1}^N}
\newcommand{\sumParticlesJExclI}{\sum_{j \neq i}}
\newcommand{\particleIndexI}{i}
\newcommand{\particleIndexJ}{j}
\newcommand{\sumLevelsOnePlus}{\sum_{l=1}^L}
\newcommand{\levelIndex}{\ensuremath{l}}
\newcommand{\gridpointIndexM}{\mathbf{m}}
\newcommand{\gridpointIndexN}{\mathbf{n}}
\newcommand{\sumGridpointsM}{\sum_{\gridpointIndexM}}
\newcommand{\sumGridpointsN}{\sum_{\gridpointIndexN}}
\newcommand{\formalVector}[1]{\ensuremath{#1}}  
\newcommand{\formalMatrix}[1]{\ensuremath{#1}}  
